\documentclass{JHEP3}
\usepackage{amsfonts,amsmath,bbm}
\usepackage{graphicx}

\newcommand{\eref}[1]{(\ref{#1})}

\newcommand{\fref}[1]{Figure~\ref{#1}}
\newcommand{\cref}[1]{Chapter~\ref{#1}}
\newcommand{\beq}{\begin{equation}}
\newcommand{\eeq}{\end{equation}}
\newcommand{\ba}{\begin{array}}
\newcommand{\ea}{\end{array}}
\newcommand{\bcenter}{\begin{center}}
\newcommand{\ecenter}{\end{center}}

\def\IC{\mathbb{C}}

\def\IF{\mathbb{F}}

\def\IGa{\relax\hbox{${\rm I}\kern-.18em\Gamma$}}

\def\IZ{\mathbb{Z}}

\def\smiley{\hbox{\large$\bigcirc$\hspace{-0.80em}\raise.2ex
\hbox{$\cdot\cdot$}\kern-.61em\lower.2ex\hbox{\scriptsize$\smile$}}\ }
\def\frowny{\hbox{\large$\bigcirc$\hspace{-0.80em}\raise.2ex
\hbox{$\cdot\cdot$}\kern-.635em\lower.2ex\hbox{\scriptsize$\frown$}}\ }

\makeatletter
\let\hangafter\@hangfrom
\makeatother

\newtheorem{thm}{Theorem}[subsection]
\newcommand{\btheorem}{\begin{thm}}
\newcommand{\etheorem}{\end{thm}}
\newtheorem{lem}[thm]{Lemma}
\newcommand{\blemma}{\begin{lem}}
\newcommand{\elemma}{\end{lem}}
\newtheorem{dfn}[thm]{Definition}
\newcommand{\bdefn}{\begin{dfn}}
\newcommand{\edefn}{\end{dfn}}
\newtheorem{cor}[thm]{Corollary}
\newcommand{\bcor}{\begin{cor}}
\newcommand{\ecor}{\end{cor}}
\def\bproof{\begin{proof}} 
\def\eproof{\end{proof}}

\newcommand{\be}{\begin{equation}}
\newcommand{\ee}{\end{equation}}
\newcommand{\bea}{\begin{eqnarray}}
\newcommand{\eea}{\end{eqnarray}}
\newcommand{\bean}{\begin{eqnarray*}}
\newcommand{\eean}{\end{eqnarray*}}
\newcommand{\bc}{\begin{center}}
\newcommand{\ec}{\end{center}}
\newcommand{\comment}[1]{}

\DeclareFontFamily{U}{rsf}{} \DeclareFontShape{U}{rsf}{m}{n}{  <5> <6> rsfs5 <7> <8> <9> rsfs7 <10-> rsfs10}{}
\DeclareMathAlphabet\Scr{U}{rsf}{m}{n} \DeclareMathAlphabet\mathbi{U}{cmr}{bx}{it}

\def\ses#1#2#3{\xymatrix@1{0 \ar[r] & #1 \ar[r] & #2 \ar[r] & #3 \ar[r] & 0}}

\def\IC{\mathbb{C}}
\def\IZ{\mathbb{Z}}

\def\be{\begin{equation}}
\def\ee{\end{equation}}

\def\CN{{\mathcal{N}}}

\preprint{
Imperial/TP/08/AH/07 \\
Bicocca-FT-09-15
\\
MIT-CTP-3977
}

\title{Brane Tilings and M2 Branes}

\author{
Amihay Hanany$^1$, David Vegh$^2$ and Alberto Zaffaroni$^3$
\\
~\\
$^1$Theoretical Physics Group, The Blackett Laboratory, Imperial College London, \\ Prince
Consort Road,
London, SW7 2AZ, U.K.\\
\vskip -0.2cm $^2$Center for Theoretical Physics,
Massachusetts Institute of Technology,\\
77 Massachusetts Avenue, Cambridge MA 02139, USA \\
\vskip -0.2cm $^3$Universit\`a di Milano-Bicocca and INFN,
sezione di Milano-Bicocca, Piazza della Scienza, 3;\\
I-20126 Milano, Italy
\\~\\
 \email{a.hanany  {\it at}  imperial.ac.uk}, \ \email{dvegh {\it at} mit.edu}, \ \email{alberto.zaffaroni  {\it at}  mib.infn.it}
 }

\abstract{Brane tilings are efficient mnemonics for Lagrangians of $\mathcal{N}=2$
Chern-Simons-matter theories. Such theories are conjectured to arise on M2-branes probing
singular toric Calabi-Yau fourfolds. In this paper, a simple modification of the Kasteleyn
technique is described which is conjectured to compute the three dimensional toric diagram of
the non-compact moduli space of a single probe. The Hilbert Series is used to compute the
spectrum of non-trivial scaling dimensions for a selected set of examples.

}

\begin{document}

\section{Introduction}

The recent work on 2+1 dimensional superconformal Chern-Simons theories which started with the
discovery of the BGL $\CN=8$  supersymmetric Chern-Simons theory \cite{Bagger:2006sk,
Bagger:2007vi,Bagger:2007jr, Gustavsson:2007vu,Distler:2008mk, VanRaamsdonk:2008ft,
Lambert:2008et} and culminated in the construction of the ABJM superconformal $\CN=6$
Chern-Simons theory \cite{Aharony:2008ug}, has shed new light on the the $AdS_4/CFT_3$
correspondence. A long-standing problem in 2+1 dimensional superconformal theories is the
identification of the gauge theory dual to an $AdS_4\times H$ supersymmetric background.  While
this problem is relatively well understood in 3+1 dimensions, it is much less understood in 2+1
dimensions. In the past, attempts to find duals have focused on Yang-Mills theories flowing in
the IR to superconformal fixed points \cite{Fabbri:1999hw, Ceresole:1999zg, Billo:2000zr,
Oh:1998qi}. It seems now that supersymmetric Chern-Simons theories can do a better job. The
$\CN=6$ ABJM theory nicely incorporates all relevant features of a dual theory for the
backgrounds with large supersymmetry, including the maximally supersymmetric case of $H=S^7$.
Other examples of superconformal Chern-Simons theories with supersymmetry $\CN=3,4,5$ have been
constructed recently \cite{Benna:2008zy, Imamura:2008nn,Hosomichi:2008jd, Hosomichi:2008jb,
Terashima:2008ba, Aharony:2008gk, Jafferis:2008qz,Fuji:2008yj}.

It is known that $AdS_4\times H$ M-theory backgrounds, with $H$ a seven dimensional
Sasaki-Einstein manifold, preserve $\CN=2$ supersymmetry\footnote{In 2+1 dimensions we can also
have $\CN=1$ supersymmetry but we do not consider this case.}
\cite{Klebanov:1998hh,Acharya:1998db, Morrison:1998cs}. The $X=C(H)$ cone over $H$ is a
Calabi-Yau four-fold and the backgrounds of interest arise as near-horizon geometries of
membranes sitting at the singular tip of the cone. We thus have a correspondence between the
infinite number of Calabi-Yau four-folds and an infinite set of superconformal theories. The
open problem is to find the explicit correspondence.  The analogous problem in 3+1 dimensions
has been solved, at least for the class of toric Calabi-Yau singularities, using Brane Tilings
\cite{Hanany:2005ve, Franco:2005rj, Franco:2007ii}. A similar proposal  for 2+1 dimensions is
based on crystals \cite{Lee:2006hw,Lee:2007kv, Kim:2007ic} but it is not as well understood as
in 3+1 dimensions. Various recent progresses have been made using Chern-Simons theories.

A 2+1 dimensional theory dual to an $AdS_4$ M-background should have various distinctive
features. In particular, the abelian moduli space should be a four-dimensional Calabi-Yau cone
$X$ and the non-abelian moduli space should be the symmetrized product of $N$ copies of $X$ (or
a modification of it). In \cite{Hanany:2008cd}, we showed how to construct infinitely many
Chern-Simons theories with these properties using tilings. For every periodic tiling of the
torus that gives rise to a consistent 3+1 dimensional superconformal gauge theory, and for every
choice of Chern-Simons parameters we constructed a Chern-Simons theory with a component of the
moduli space which is the symmetric product of a Calabi-Yau four-fold. Each tiling gives rise
therefore to an infinite family of Calabi-Yau four-folds depending on the integer Chern-Simons
parameters. As independently noticed by various authors \cite{Hanany:2008cd, Jafferis:2008qz,
Martelli:2008si}, the abelian moduli space of quiver Chern-Simons theory is naturally
four-dimensional. In \cite{Hanany:2008cd} we demonstrated that the mesonic abelian moduli space
of 2+1 dimensional quivers arising from tilings is always a toric Calabi-Yau cone and that the
full mesonic moduli space is generically a symmetric product\footnote{See \cite{Martelli:2008si}
for an independent analysis.}.

In this paper we continue our analysis of the class of Chern-Simons theories obtained from
tilings following two important directions.

First of all, it is important to find an efficient forward algorithm, {\it i.e.} a prescription
for determining the toric diagram of the Calabi-Yau four-fold. This problem was addressed in
\cite{Hanany:2008cd} by explicitly looking at the supersymmetric moduli space of vacua as a set
of solution of F and D terms. The recent advances in understanding the space of solutions of the
F-terms for a toric quiver \cite{Forcella:2008bb,Forcella:2008eh} - the {\it master space} as we
dubbed it - allows to work out easily many examples from this perspective \cite{Hanany:2008cd}.
This forward algorithm, however, become increasingly cumbersome as the size of the tiling is
increased. We need therefore to find a fast forward algorithm, similar to that existing in 3+1
dimensions \cite{Franco:2005rj}. We indeed show that, a mild modification of the 3+1 dimensional
algorithm gives an efficient fast forward algorithm also in 2+1 dimensions: it is based, as in
3+1 dimensions, on the concepts of perfect matchings and of the Kasteleyn matrix.

Secondly, it is important to study the quantum properties of the Chern-Simons theory and its
spectrum of conformal dimensions from the dual supergravity perspective. The Chern-Simons
theories are expected to flow to IR fixed points. The superpotential is not always quartic and
therefore the R-charge of the fields and the dimension of gauge invariant operators are quantum
corrected. There are many abelian global symmetries in the quiver that mix with the R-symmetry.
There is a notion of exact R-symmetry at the IR fixed point, which is the one sitting in the
superconformal algebra, but it is extremely difficult to find it. At the moment of this writing,
the many efficient 3+1 dimensional tools for studying superconformal theories are not available
in 2+1 dimensions. In particular we have nothing similar to a-maximization
\cite{intriligator:2003jj} to predict the exact R-symmetry. However, the spectrum of conformal
dimensions of a Chern-Simons theory dual to an $AdS_4\times H$ background can be predicted from
supergravity. We use geometrical methods to find the exact R-charges of the dual Chern-Simons
theory -- in particular minimization of the volume functional and computation of volumes
\cite{Martelli:2006yb}. In 3+1 dimensions these tools are the geometrical counterpart of
a-maximization \cite{Martelli:2005tp, Butti:2005vn,Butti:2005ps}. We show, in all our examples,
that the supergravity results for the mesonic spectrum are consistent with the Chern-Simons
expectations and extremely similar in structure to the results in 3+1 dimensions. We also look
at baryons that, as familiar in the AdS/CFT correspondence, correspond to wrapped branes and we
point out some puzzles. We notice that the consistency of the duality requires the understanding
of the absence of certain supersymmetric states in the gravitational dual or, equivalently, the
presence of new states in the Chern-Simons theory.

It is important to know what are the prediction of supergravity. Although we cannot compare the
result with exact field theory computations,  we can predict the full spectrum of dimensions of
the dual Chern-Simons theory and we can make various checks of the overall consistency of the
construction. It is not obvious for example that every tiling gives rise to a consistent
superconformal theory. Moreover, it seems that various different tilings, and more generally
various  different quivers, give rise to the same Calabi-Yau four-fold. The knowledge of the
spectrum may help in understanding whether these models are equivalent or which is the best
candidate for a duality.

We would like to stress  the important role played by the master space of the 3+1 dimensional
model \cite{Forcella:2008bb}.  The master space encodes all the properties of the Chern-Simons
theory in a beautiful way. In particular, hidden symmetries in the master space of some tilings
\cite{Forcella:2008bb}, that are still mysterious from the point of view of 3+1 dimensions,
reveal their role in the 2+1 dimensional theories.

The paper is organized as follows. In Section 2 we review the construction of Chern-Simons
theories from tilings \cite{Hanany:2008cd}. In Section 3 we explain the fast forward algorithm
in terms of the Kasteleyn matrix and the procedure for determining the toric data of the
four-dimensional Calabi-Yau singularity. In Section 4 we discuss the physical meaning of the
Hilbert series. We also distinguish between mesonic and baryonic spectra. In Section 5 we
elaborate on many examples. The paper ends with conclusions and comments. The first Appendix
contains technical results on the computation of Hilbert series that are used in the main text.
The second Appendix describes some observations about brane crystals and some speculations about
a possible (Seiberg) dual theory to the ABJM theory at level one.

While finishing this work, two papers appeared that have partial overlap with Section 3
\cite{Ueda:2008hx, Imamura:2008qs} and that present an algorithm for computing the toric data of
the Calabi-Yau four-fold. These algorithms seem to be equivalent to the fast forward algorithm
presented here.

\section{Brane tilings and M2 branes}

Recall from \cite{Hanany:2008cd} that for every (consistent) periodic, bipartite, two
dimensional tiling of the plane we can construct a ${\cal N} =2$ (4 supercharges) CS theory in
2+1 dimensions whose abelian moduli space is a toric Calabi-Yau four-fold.

The rules for writing down the 2+1 dimensional theory follow the rules set out for the 3+1
dimensional theory \cite{Hanany:2005ve,Franco:2005rj}. Every face is a $U(N)$  gauge group and
every edge is a chiral superfield transforming in a bifundamental representation of the two
gauge groups it separates with orientation defined by the bipartite structure of the tiling. By
convention one can pick an (incoming) outgoing arrow to correspond to an (anti)-fundamental
representation, respectively.  Every vertex in the tiling contributes a term in the
superpotential given by the products of all the fields that meet at the vertex\footnote{The
coefficients in the superpotential are not encoded in the tilings.}, with a positive sign for
white vertices and a negative sign for black ones. This set of rules corresponds to a
Hanany-Witten construction of the theory, where the faces of the tiling are D4-branes bounded by
NS-branes and chiral superfields arise from open strings connecting adjacent faces. As the
theory flows to the IR, one should lift the configuration to M-theory where it becomes a theory
of M2 branes.

\begin{figure}[ht]
\begin{center}
  \includegraphics[totalheight=4.0cm]{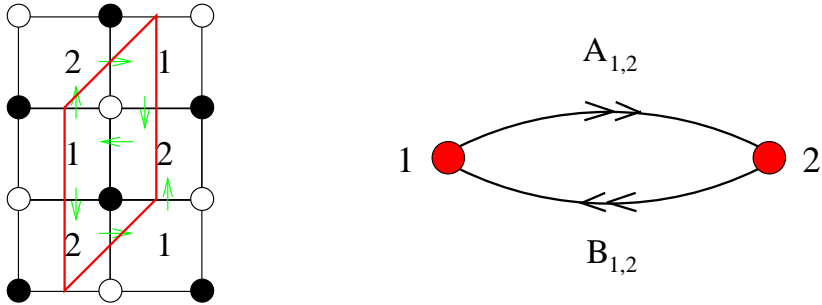}
  \caption{(i) Brane tiling for $\IC^4$ (and $\IC^4/\IZ_k$). The fundamental domain is shown in red. The green
  arrows indicate the direction of the bifundamental fields based on the convention that the black node is on the left-hand side.  \ (ii) The corresponding quiver.}
  \label{coni1}
\end{center}
\end{figure}

An example (in fact the simplest) is the ABJM theory with ${\cal N}=6$ supersymmetry
\cite{Aharony:2008ug}. It is obtained from the tiling that gives rise to the conifold theory in
3+1 dimensions. This is shown in Figure \ref{coni1}(i) with the quiver in Figure
\ref{coni1}(ii). There are two gauge groups, fields $A_i,B_i,\, i=1,2$ transforming in the
$(N,\bar N)$ and $(\bar N, N)$ representation of the gauge group, respectively, and interacting
with the superpotential
\beq W= A_1 B_1 A_2 B_2 - A_1 B_2 A_2 B_1 .
\label{conisupo}\eeq

In 3+1 dimensions, we would introduce standard kinetic terms for every gauge group and we would
obtain a superconformal quiver gauge theory with an abelian moduli space that is a toric Calabi-Yau
three-fold.

In 2+1 dimensions, we are not introducing kinetic terms for the gauge fields but instead CS
interactions. For each edge $E_i$ we add an $\CN=2$ preserving Chern-Simons interaction with the
following rule: add an integer CS coefficient $k_i$ and $-k_i$ to the adjacent gauge groups
connected by the edge. Call $k_a$ the resulting CS coefficient for the $a$-th gauge group. The
supersymmetric vacua are obtained as usual as solutions of the F- and D-term constraints.
Consider for the moment the abelian theory. As extensively discussed in \cite{Hanany:2008cd},
the F-term constraints are given by the vanishing of all derivatives of the superpotential, as
in 3+1 dimensions, and the D-terms constraints can be summarized by the following equations
\beq
\mu_a(X) = 4 k_a \sigma
\eeq
where $\mu_a(X)$ is the moment map for the action of the $a$-th group (the 3+1 dimensional
D-term) and $\sigma$ is an auxiliary field in the vector supermultiplets.  Since $\sum_a k_a=0$
and $\sum_a \mu_a(X)=0$ by construction, one of these equations is redundant; the overall $U(1)$
does not enter in the supersymmetric vacua conditions, as usual. Moreover, any other linear
combination of gauge groups with coefficient $m_a$ orthogonal to the CS parameters $\sum_a k_a
m_a=0$ has a vanishing moment map. We are thus imposing $g-2$ D-term constraints, where $g$ is
the number of gauge groups. As in 3+1 dimensions, we can impose simultaneously the D-term
constraints and the corresponding $U(1)$ gauge transformations by modding out by the
complexified gauge group. In other words, we are modding by the $g-2$ dimensional subspace of
the natural $(\mathbb{C}^*)^{g}$ action that is in the kernel of the matrix
\beq
{\scriptsize
C=\left( \begin{matrix}
   1 & 1 & \cdot\cdot\cdot & 1 & 1 \cr
   k_1 & k_2 & \cdot\cdot\cdot & k_{g-1} & k_g \cr
  \end{matrix} \right) \ .
}
\eeq
We do not need to divide by the remaining $U(1)$ gauge field, since the last equation just
determines the value of the auxiliary field $\sigma$. However, through its CS coupling with the
overall gauge field, it leaves a discrete symmetry $\mathbb{Z}_k$, where $k=\gcd (\{k_a\})$.

It is then easy to show that the resulting moduli space is a toric Calabi-Yau four-fold. The
solution of the F-term constraints is a $g+2$ dimensional toric Calabi-Yau variety called {\bf
the master space} \cite{Forcella:2008bb,Forcella:2008eh}.  All the complexified gauge groups,
except the overall one, act on this variety as non-trivial $\mathbb{C}^*$ toric actions. By
modding out by the $(\mathbb{C}^*)^{g-2}$ kernel of $C$ we obtain, as promised, a Calabi-Yau
four-fold\footnote{The resulting variety is Calabi-Yau because the vectors of charges in
$\mathbb{C}^{g-2}$ are traceless by construction.}.  The moduli space is obtained by a further
modding by the remaining discrete $\mathbb{Z}_k$ symmetry. This moduli space is interpreted as
the transverse space to one M2-brane in M-theory which probes the CY four-fold.

As shown in \cite{Hanany:2008cd}, the non-abelian mesonic moduli space is generically the $N$-fold
symmetrized product of the Calabi-Yau four-fold.

In this paper, following the notation introduced in \cite{Hanany:2008cd},  we use the Calabi-Yau
three-fold $Y$ appearing in the 3+1 dimensional theory to identify the 2+1 dimensional theory. A
Chern-Simons theory associated with the tiling for the Calabi-Yau three-fold $Y$ with
Chern-Simons parameters $k_a$ is denoted $\widetilde{Y}_{\{ k_a\}}$.

\section{The Calabi-Yau four-fold}

The natural and immediate question is how to determine the toric data of the Calabi-Yau
singularity. We emphasize that, in principle,  all the information about the variety is encoded
in the computation of the moduli space as a solution of F- and D-term constraints given in the
previous section as a symplectic quotient of the master space. The toric properties of the
master space have been studied in \cite{Butti:2007jv,Forcella:2008bb,Forcella:2008eh} and can be
used to extract the toric data of the four-fold. In particular we always have a forward
algorithm based on finding the kernel of the matrix of charges in the symplectic quotient
description of the moduli space. The master space is however quite big and the previous analysis
is done on a case by case basis. We therefore need to look for a more efficient algorithm to
compute the toric diagram of the four-fold from the tiling data. Recall that, in 3+1 dimensions,
the analogous algorithm is provided by the Kasteleyn matrix \cite{Hanany:2005ve,Franco:2005rj}.
We next show that a simple modification of the prescription works also in 2+1 dimensions.

We first explain the prescription using the simple ABJM theory as an example.

\subsection{Toric data of the ABJM theory} \label{conifold}

The tiling and quiver of this theory are depicted in \fref{coni1}, and the superpotential is
given in \eref{conisupo}. We set the Chern-Simons levels to $(k,-k)$. In the abelian theory the
F-terms are trivial and there are no D-terms to be imposed. The remaining discrete symmetry is
given by a $\mathbb{Z}_k$ subgroup of the gauge group. The moduli space is then
$\mathbb{C}^4/\mathbb{Z}_k$.

\begin{figure}[ht]
\begin{center}
  \includegraphics[totalheight=6cm]{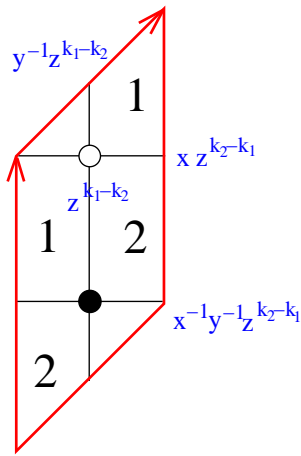}
  \caption{Fundamental cell of the $\IC^4$ brane tiling. The weights of the four edges are shown in blue. }
  \label{coni2}
\end{center}
\end{figure}

The fundamental cell of the periodic graph is shown in \fref{coni2}. The toric diagram of the
moduli space is computed by taking the permanent\footnote{The permanent is similar to the
determinant: the signatures of the permutations are not taken into account and all terms come
with a $+$ sign. One can also use the determinant but then certain signs must be introduced
\cite{Hanany:2005ve,Franco:2005rj}.} of a certain adjacency matrix of the graph. The rows of
this matrix correspond to black nodes and the columns correspond to white nodes. An element of
the matrix therefore selects a black and a white vertex in the graph. If there is no connection
between them, then the element is zero. If there is an edge between them, then the matrix
element is one. If the edge crosses the boundaries of the fundamental domain, then one needs to
multiply by a corresponding weight $x$ or $y$ (or $x^{-1}$ or $y^{-1}$, depending on the
orientation of the edge). The resulting matrix is called the Kasteleyn matrix.

The permanent of the matrix is a Laurent polynomial of two variables $x$ and $y$. For 3+1
dimensional theories, the Newton polygon of this polynomial gives the toric diagram of the
Calabi-Yau threefold. This polygon is constructed by taking the convex hull of a set of points
in an integer lattice. These points are in one-to-one correspondence with the terms in the
polynomial and their position in the lattice is given by the exponents of the $x$ and $y$
weights. The resulting two-dimensional diagram is the toric diagram of the threefold moduli
space (recall that the triviality of the canonical class restricts the endpoints of the vectors
to be on a plane).

In order to avoid dividing by the overall $U(1)$, we modify this simple algorithm by introducing
a third weight $z$. This makes it possible to obtain three dimensional toric diagrams which in
turn define Calabi-Yau fourfolds. We give a bifundamental field between gauge groups $a$ and $b$
a weight $z^{k_a-k_b}$. Here $k_a$ and $k_b$ are the Chern-Simons levels for the groups. The
full set of weights for the ABJM theory is shown in \fref{coni2}.

\begin{figure}[ht]
\begin{center}
  \hskip -10cm
  \includegraphics[totalheight=1.5cm]{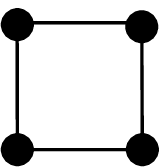}
  \vskip -2.2cm
  \includegraphics[totalheight=2.5cm]{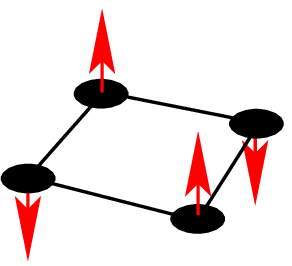}
  \vskip -2.6cm
  \hskip 10cm
  \includegraphics[totalheight=3.0cm]{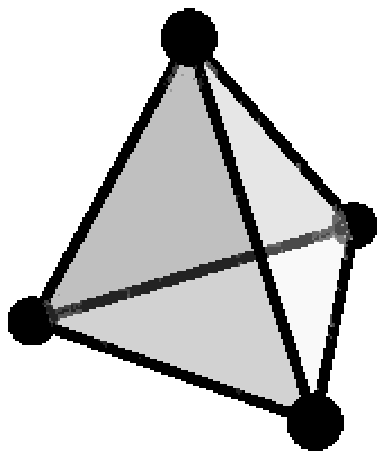}
  \caption{(i) Toric diagram for the conifold. \ (ii) Introducing the level translates the points as shown by the arrows.
  \ (iii) The resulting toric diagram is that of $\IC^4$.}
  \label{coni_toric}
\end{center}
\end{figure}

Since the fundamental cell of the ABJM theory contains only one black and one white node, the
Kasteleyn matrix is $1\times 1$,
\be
  K = \frac{\tilde z}{y} + \tilde z + \frac{x}{\tilde z} + \frac{1}{xy\tilde z}
\ee
where $\tilde z = z^{k_1-k_2} = z^{2k}$. By setting $\tilde z=1$, the 2d Newton polygon gives
the conifold toric diagram which is a square up to an $SL(2, \IZ)$ transformation. If we keep
$\tilde z$, then we obtain a 3d diagram which is shown in \fref{coni_toric}. This is a
tetrahedron and it is the toric diagram for $\IC^4$ when $k=1$. As the level $k$ increases, the
tetrahedron gets stretched in the third dimension. In toric geometry, such rescaling is
equivalent to orbifolding and thus the Kasteleyn algorithm reproduces the $\IC^4 / \IZ_k$ ABJM
moduli space.

The same algorithm works for more complicated tilings as is demonstrated in many explicit
examples in this paper. In the general case, however, an issue arises: after proper
normalization, the resulting toric diagrams are typically not at integer points in the 3d
lattice. This makes it sometimes harder to identify the moduli spaces. This problem may be
addressed by an appropriate ``gauge transformation'' of the weights on the tiling edges such
that they give rise to integer toric diagrams.

\subsection{Kasteleyn matrix and perfect matchings} \label{kasteleyn}

Before proceeding, let us discuss the useful concept of perfect matching. We briefly review it
because it is used in the following.  The reader is referred to
\cite{Hanany:2005ve,Franco:2005rj,Kennaway:2007tq} for a comprehensive introduction. A perfect
matching is a subset of edges in the tiling, or equivalently a subset of elementary fields,
that covers each vertex exactly once. The permanent of the Kasteleyn matrix is precisely a
generating function for perfect matchings. Each perfect matching is therefore associated with an
integer point in the toric diagram. We can see this by writing a refined Kasteleyn matrix by
inserting also the field corresponding to each edge. For the ABJM theory we have for example
\be
K = B_1 \frac{\tilde z}{y} + B_2 \tilde z + A_1 \frac{x}{\tilde z} + A_2 \frac{1}{xy\tilde z}
\ee
Each term corresponds to one of the four perfect matchings, $p_1=A_1,p_2=A_2,p_3=B_1,p_4=B_2$.
The ABJM theory is particularly simple since the Kasteleyn matrix is one by one. In a more
general situation, each monomial in ${\rm perm}\, K$ would be the product of the fields
belonging to a given perfect matching: $\prod_{X_i\in p_\alpha} X_i$. We discuss many explicit
examples below.

The importance of perfect matchings comes from the fact that they generate the master space
\cite{Forcella:2008bb}. In fact, as shown in \cite{Hanany:2005ve,Franco:2005rj}, the perfect
matchings $p_\alpha$ parameterize the solutions of the F-term conditions through the formula
\begin{equation}
X_i =\prod_{\alpha=1}^c p_\alpha^{P_{i \alpha}} \ .
\label{pmpar}
\end{equation}
Here the matrix $P$ contains entries which are either $0$ or $1$, encoding whether a field $X_i$ in the quiver is in the perfect matching $p_\alpha$:
\begin{equation}
P_{i \alpha} =
\begin{cases} 1 & \text{if $X_i \in p_\alpha$,} \\
0 &\text{if $X_i \not \in p_ \alpha $.}
\end{cases}
\end{equation}
For the conifold theory the situation is trivial, since there is a one to one correspondence between
fields and perfect matchings. The master space is just $\mathbb{C}^4$. In general, there are
more perfect matchings than fields, and the parametrization (\ref{pmpar}) is defined only modulo
some redefinition of the $p_\alpha$. We get a description of the master space as a Gauged Linear
Sigma Model (GLSM), or symplectic quotient: $\mathbb{C}^c/(\mathbb{C}^*)^{c- 2 -g}$. The vectors
of charges appearing in this description can be read off from the tiling as the linear relations
satisfied by the perfect matchings $\sum_\alpha Q_\alpha p_\alpha=0$ considered as formal linear
combinations of edges.

We emphasize the conceptual importance of the perfect matchings. In 3+1 dimensions, they
correspond to the integral points in the toric diagram including the internal ones and including
multiplicities (internal points have multiplicity greater than 1 and external points have
multiplicity 1. Points on the boundary of the toric diagram are binomial coefficients
\cite{Feng:2002zw}). The perfect matchings corresponding to the external points of the 2d toric
diagram, with multiplicity one, can be used to parametrize all the non-anomalous charges of the
3+1 dimensional theory \cite{Franco:2005sm,Benvenuti:2005ja,Butti:2005vn}. In 2+1 dimensions,
the 3d toric diagram is a split version of the 2d one, where some multiplicities have been
lifted, keeping the total number of perfect matchings fixed. The perfect matchings still
correspond to integer points in the 3d toric diagram. As we show, at least in all the examples
considered in this paper, the external perfect matchings of the 3d toric diagram can be used to
parametrize the $g+2$ charges of the 2+1 dimensional theory. Finally, if we grow in dimensions
with the purpose of studying the master space, we discover that all multiplicities have been
lifted and the perfect matchings correspond exactly to the external point of the toric diagram
for the coherent component of the master space that is a $g-2$ dimensional Calabi-Yau
singularity \cite{Forcella:2008bb}.

It is also important to note that the total number of perfect matchings is preserved by our
construction. An important corollary is that only 3d toric diagrams which have a 2d projection
such that the resulting 2d toric diagram corresponds to a consistent 3+1 dimensional
theory\footnote{See \cite{Gulotta:2008ef} for the most updated discussion on consistency of 2d
tilings as giving rise to 3+1 dimensional theories.} are represented by a 2d tiling.

\section{The Hilbert series}\label{hilbert}

In this paper we describe the properties of the Calabi-Yau four-fold by means of its Hilbert
series. Recall that the Hilbert Series has many equivalent interpretations:
\begin{itemize}
\item
Mathematically, it is the generating function for holomorphic functions on the Calabi-Yau
$X=C(H)$.
\item
Physically, from the field theory point of view, it is the partition function that counts chiral
mesonic operators for the theory on one membrane. Thanks to the beautiful structure of the
moduli space, which is a symmetric product, it is also the partition function for single trace
chiral mesonic operators for large $N$ \cite{Benvenuti:2006qr}.
\item
Finally, from the point of view  of the compactification on $AdS_4\times H$, it is the
generating function for KK chiral multiplets \cite{Hanany:2008qc}. In fact every holomorphic
function on $X$ descends to to an eigenvector of the Laplacian on the base $H$.
\end{itemize}

When the four-fold ${X}$ is toric we can refine the Hilbert series with four weights $t_i$
corresponding to 4 global $U(1)$ symmetries, some of which may be subgroups of a bigger
non-abelian symmetry group. The meaning of the refined Hilbert series is to count operators
according to their global and R-charges. One important geometrical property of the Hilbert
Series is that, for $t_i\rightarrow 1$, it computes the volume of the base $H$
\cite{Martelli:2006yb}. In fact, setting $t_i = e^{-\mu b_i}$ we have for $\mu\rightarrow 1$
\beq
g(t_i; { X}) \sim  \frac { {\rm Vol}(b_i)}{\mu^4} +\cdot\cdot\cdot
\eeq
where the numerator is the volume of the family of Sasaki manifold with Reeb vector
$R=\{b_1,b_2,b_3,b_4\}$.

At the superconformal fixed point there is a notion of the exact R-symmetry, which sits in the
superconformal algebra. The exact R-symmetry corresponds to the Reeb vector that defines a
Calabi-Yau metric on the cone. As shown by \cite{Martelli:2005tp}, the Reeb vector can be found
by minimizing the function ${\rm Vol}(b_i)$. The minimization is done on the three parameter set
of $b_i$ that give R-charge 2 to the holomorphic top form of the CY. The minimization, in turn,
determines the exact R-symmetry. From the Hilbert series we can then compute the spectrum of
dimensions of all mesonic operators, which agrees with the KK computation. For a $U(N)$ theory
this is the complete spectrum of chiral operators. The R-charges of mesonic operators can be
read from minimization.

Let us see the Hilbert series and minimization in action for the simple ABJM theory.
There are 4 perfect matchings in correspondence with 4 quiver fields. One can assign to each
point in the toric diagram a perfect matching, or alternatively, to each point in the toric
diagram a quiver field. The moduli space is $\IC^4$, parametrized by four coordinates, $A_1,
A_2, B_1, B_2$. The global symmetry is $U(4)$ with rank 4 and one can assign four fugacities
$t_i, i=1\ldots 4$, each counting the number of fields of type $i$. The refined Hilbert series
is particularly simple since the moduli space is freely generated by four variables
\beq
g \left (t_i, \tilde{{\cal C}}_{\{1, -1\}} \right ) = \frac{1}{(1-t_1)(1-t_2)(1-t_3)(1-t_4)}
\eeq
To compute R-charges we introduce chemical potentials $t_i=\exp(-\mu b_i)$ and take the limit
$\mu\rightarrow 0$, with the coefficient of the most singular piece,
\beq
\lim_{\mu\rightarrow0} \mu^4 g \left (e^{-\mu b_i}, \tilde{{\cal C}}_{\{1, -1\}} \right ) =
\frac{1}{b_1 b_2 b_3 b_4},
\eeq
and impose the CY condition $b_1 + b_2 + b_3 + b_4 = 2$. This condition comes from the fact that
we require the top holomorphic form $dt_1\wedge dt_2\wedge dt_3\wedge dt_4$ to have R-charge
$2$. Minimizing this expression we find $b_i=1/2, i=1\ldots4$, leading to R-charges $1/2$ per
each field which is the canonical dimension for a scalar field in 2+1 dimensions. Since the
global symmetry has a non-abelian factor we expect no contributions from the corresponding
chemical potentials. We set $t_1 = t x_1, t_2 = t x_2 / x_1, t_3 = t x_3 / x_2, t_4 = t / x_3$,
where $t$ corresponds to the $U(1)_R$ symmetry, and $x_1, x_2, x_3$ are weights of the
non-abelian $SU(4)$ symmetry. We further set $t = e^{-\mu b}$ and the CY condition takes the
form $4b =2$, which immediately gives the desired answer $b =1/2$.

In more complicated examples, the Calabi-Yau four-fold will be given by a symplectic quotient
$\mathbb{C}^d/(\mathbb{C}^*)^{d-4}$ in some ambient space of dimension $d$. We can then use an
integral Molien formula to compute the Hilbert series
\beq \oint \prod_{i=1}^{d-4} \frac{dz_i}{2\pi i z_i} \frac{1}{\prod_{a=1}^{d} (1 - \tilde t_a Z_a )} \eeq
where $\tilde t_a=\tilde t_a (t_i)$ is a convenient parametrization of the coordinates in
$\mathbb{C}^d$ in terms of the four toric charges and $Z_a=Z_a(z_i)$ denotes the monomial weight
of the $a$-th coordinate in terms of the $(\mathbb{C}^*)^{d-4}$ group. We refer to Appendix
\ref{prehS} for a detailed explanation of the Molien formula.

We always have such a description of the Calabi-Yau four-fold as a symplectic quotient. One
description is familiar from toric geometry \cite{fulton} and it is obtained from the toric data
provided by the Kasteleyn matrix. Another description follows from the explicit construction of
the moduli space in field theory: the CY four-fold can be written as a symplectic quotient of
the coherent component of the master space of dimension $g+2$, which  is itself a symplectic
quotient of the space of perfect matchings \cite{Forcella:2008bb,Forcella:2008eh}. Since the
number of perfect matching is large this description is sometimes cumbersome, even if more
physical than the purely geometric one. In some lucky cases, we will be able to write the master
space or the moduli space as a set of algebraic equations describing a complete intersection
variety; in all these cases the computation of the Hilbert series simplifies.

\subsection{Baryonic charges}

Now recall that the AdS/CFT correspondence applies to $SU(N)$ theories. This means that the
number of global symmetries of our theories is larger than four; it is actually $g+2$, where $g$
is the number of gauge groups\footnote{ We see that the number of global symmetries that can mix
with the R-charge is greater than the analogous number in the 3+1 dimensional theory associated
with the same tiling, which is the number of external points of the 2d toric diagram minus
three. The reason is that in 3+1 dimensions some global symmetries are anomalous while in 2+1
dimensions there is no anomaly.}.  For the ABJM theory $g+2=4$ is the number of toric symmetry
of the four-fold and there is no baryonic charge, but, in a general tiling, $g+2 >4$ and there
is plenty of baryonic charges. The mesonic spectrum does not depend on the new charges, but the
baryonic spectrum does. Using only the Hilbert series and minimization we cannot determine the
R-charges of all fields.

This is usually solved by looking explicitly at baryons. Recall that in the AdS/CFT
correspondence baryons appear as wrapped branes. In 2+1 dimensions, a simple scaling argument
says that, in order to have objects with dimensions proportional to $N$, we need to wrap M5
branes on five-cycles in $H$. The R-charge of a five-brane wrapped on the cycle $\Sigma_5$ in
$H$ is given as a normalized volume by the familiar formula \cite{Gubser:1998fp}
\beq  \label{Rc} R\equiv \Delta = \frac{\pi N}{6} \frac{{\rm Vol}(\Sigma_5)}{{\rm Vol}(H)} .\eeq
We refer to the particular ratio of volumes appearing in the previous formula as {\it normalized
volume} in the rest of the paper. As usual in toric geometry, the five-cycles $\Sigma_5$ are associated with divisors $D_a$ in ${ X}$  and with the external points $v_a$ in the 3d toric diagram \cite{fulton}. The normalized volumes can be
easily computed using geometrical methods for all toric Calabi-Yau four-folds ${ X}$. The technical
results on computation of volumes are given in Appendix \ref{prehS}.

As is well known, an analogous computation in 3+1 dimensions gives the exact R-charges of all
elementary fields. The reason is that all wrapped branes appear as baryons made with elementary
fields. There is indeed a set of fields in the 3+1 dimensional quiver that is directly
associated with the divisors $D_a$ and whose R-charge can be computed by using normalized
volumes. All the other elementary fields are associated with integer linear combinations of
divisors and their R-charge is known by the additive property of the charge
\cite{Franco:2005sm,Benvenuti:2005ja,Butti:2005vn}.

We analyze in detail many examples in 2+1 dimensions. We note that in all our examples the
elementary fields can be set to be in correspondence with linear combinations of divisors, or
equivalently linear combinations of five-cycles, although not all such combinations appear. The
value of  the exact R-charges corresponding to a five-brane wrapped on a divisor base is given
by formula (\ref{Rc}).  Unfortunately, there is no efficient tool in 2+1 dimensions to compute
the exact R-charge at the fixed point, so the  comparison with field theory can be only
qualitative. We also note that not all the wrapped branes appear as elementary fields; the
existence of other states in the CS theory remains to be properly understood.

We conclude this section with a technical remark. In all our examples the external perfect
matchings in the 3d toric diagram can be used to parametrize the $g+2$ global charges, exactly
as in 3+1 dimensions the external perfect matchings of the 2d toric diagram parametrize the non
anomalous global charges. For the generic case of a regular 2d toric diagram which becomes a
regular 3d toric diagram, we expect exactly $g+2$ external points. Recall indeed that the number
of gauge groups is given by the area of the 2d toric diagram that, by Pick's theorem
\cite{fulton}, is $g=n + 2i -2 $ where $n$ is the perimeter of the toric diagram (the number of
integer points on the boundary) and $i$ is the number of integer internal points. In a regular
2d diagram without integer points on the sides, the external points have multiplicity one and
they remains external points in the 3d diagram. An internal point has multiplicity, but in the
3d diagram only two of the split points appear as external. The counting now generically
reproduce $g+2=n+2i$. The external points, and the corresponding perfect matchings can be then
used to parametrize the global charges. More care should be used in cases where there are points
on the sides, and this will be analyzed case by case in the following examples.

\section{Examples}
\subsection{The $\widetilde{\IC^2 / \IZ_2 \times \IC}$ theory} \label{c2z2sec}

This example is special since it is the simplest model with a spectrum of non-trivial scaling
dimensions. It has two gauge groups just like the modified conifold theory of Section
\ref{conifold}, and 6 fields -- slightly more than the 4 of Section \ref{conifold} but more
crucially it has a cubic superpotential. This is to be contrasted with the modified conifold
theory which has a quartic superpotential.

Let us re-consider the modified conifold case briefly. Symmetry reasoning leads to the natural
scaling dimension $1/2$ for all basic fields consistent with the value for a free field in 2+1
dimensions. This argument implies that the simplest interaction term of the superpotential in
2+1 dimensions is quartic, different from the simplest interaction term in 3+1 dimensions which
is cubic. Indeed the free field scaling dimension in 3+1 dimensions is 1 and to get a
superpotential with an interaction term of dimension 3 we would take a cubic term. By using this
simple scaling argument we learn that any interaction term in 2+1 dimensions which is not
quartic must lead to non-trivial scaling dimensions as it is not possible to assign scaling 1/2
to all fields and still get a total scaling of 2. At least one of the fields must exhibit some
strong coupling effect. The simplest interaction term with these property is therefore the cubic
term. We therefore crown the current example as the simplest model with a spectrum of
non-trivial scaling dimensions for chiral operators. Below we will list the scaling dimensions
and indicate a method for computing them using simple tools in toric geometry and tools from the
theory of two dimensional tilings.

\begin{figure}[ht]
\begin{center}
  \includegraphics[totalheight=3.5cm]{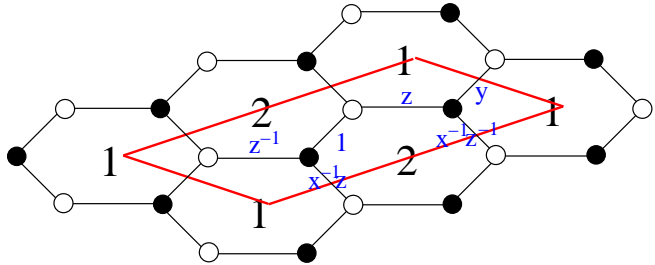}
  \caption{Brane tiling for $\mathcal{C} \times \IC$ with edge weights around the two black nodes.}
  \label{cc1fig}
\end{center}
\end{figure}

Figure \ref{cc1fig} shows the tiling for this theory. The theory has two gauge groups, two
adjoint fields $\Phi_i$ and four chiral fields $A_i,B_i,\, i=1,2$ transforming in the $(N,\bar
N)$ and $(\bar N, N)$ representation of the gauge group, respectively, and interacting with the
superpotential
\beq W=  \Phi_1(A_1 B_2 - A_2 B_1) +\Phi_2 ( B_2 A_1 - B_1 A_2)  \eeq

The moduli space of the 2+1 dimensional theory is the conifold times the complex plane
\cite{Hanany:2008cd}. Since the number of gauge group is $2$ there is no D-term to be imposed
and the moduli space of the CS theory coincides with the coherent component of the master space,
which is indeed four-dimensional.  We note that the master space is reducible. On the coherent
component of the moduli space the relation $\phi_1=\phi_2$ is valid
\cite{Hanany:2006uc,Forcella:2007wk,Forcella:2008bb} and we will generically denote the
independent adjoint field as $\phi$. A similar remark applies to other orbifold examples in this
paper.

To compute the toric diagram for CS levels $\{1,-1\}$ we write the Kasteleyn matrix,
\be
  K =   \left(
\begin{array}{cc}
  z^{-1}+ z x^{-1} & 1  \\
  y & z+z^{-1} x^{-1}
\end{array}
\right)
\ee
and the permanent gives
\be
  \textrm{perm} \, K = 1+ \frac{1}{z^2 x} + \frac{z^2}{x} + \frac{1}{x^2} + y .
\ee

\begin{figure}[ht]
\begin{center}
  \vskip 0.8cm
  \hskip -4cm
  \includegraphics[totalheight=1.5cm]{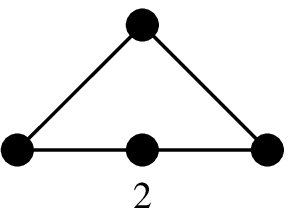}
  \vskip -2.5cm
  \hskip 6cm
  \includegraphics[totalheight=3.0cm]{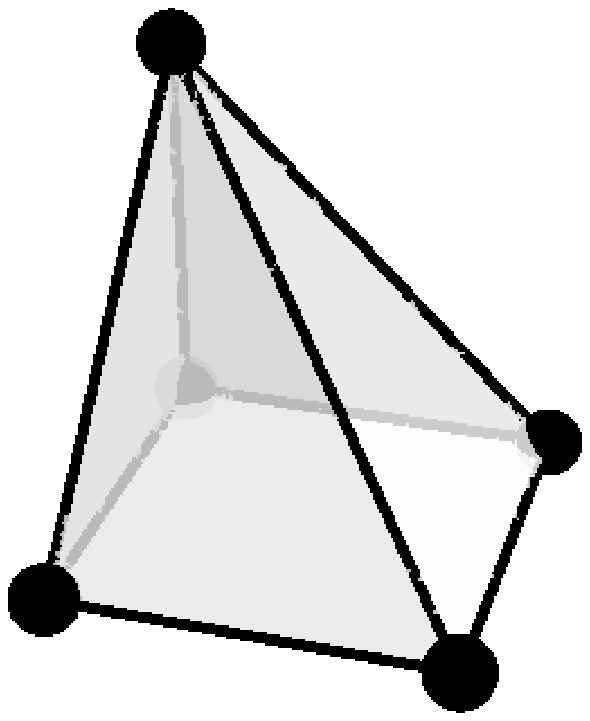}
  \caption{(i) The 2d toric diagram for $\IC_2 / \IZ_2 \times \IC$, denoted below by ${\cal T}_2$. \ (ii) The 3d toric diagram for $\mathcal{C} \times \IC$, denoted below by ${\cal T}_3$. The internal point of multiplicity 2 in ${\cal T}_2$ splits into two external points in ${\cal T}_3$.}
  \label{cctd3d}
\end{center}
\end{figure}

\begin{figure}[ht]
\begin{center}
  \includegraphics[totalheight=2.3cm]{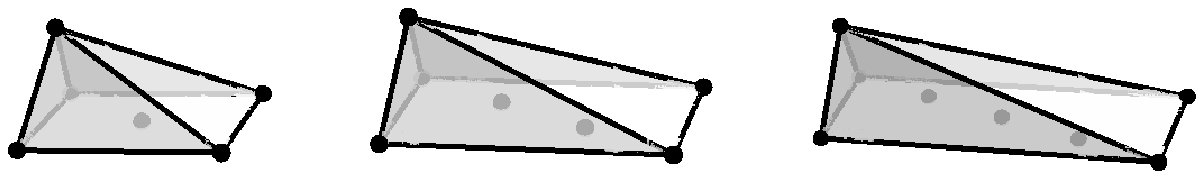}
  \caption{The 3d toric diagram for $\mathcal{C} \times \IC$ orbifolds, with $\IZ_2$, $\IZ_3$ and $\IZ_4$, respectively}
  \label{cczkfig}
\end{center}
\end{figure}

This result shows that the double point of the original toric diagram of $\IC_2 / \IZ_2$ gets
separated in the third dimension, as depicted in \fref{cctd3d}. However, in order to get the
minimal volume we need a redefinition $z \rightarrow z^{1/4}$. Without this we would get a toric
diagram that has internal points and corresponds to an orbifold of the $\{1,-1\}$ theory.
Examples of the toric diagram for CS levels $\{2,-2\}, \{3,-3\}, \{4,-4\}$, respectively, are
shown in \fref{cczkfig}.

Next, we look at quiver fields and perfect matchings in order to prepare the ground for the
scaling dimensions. For this example there are 6 quiver fields $\phi_1, A_1, B_1, \phi_2, A_2,
B_2$. We first use a symmetry argument to compute the scaling dimensions, before making a
detailed analysis. The $\IC$ part of the moduli space is not changing in this problem and
therefore we expect the fields parametrizing it, $\phi$, to have the trivial scaling dimension
$1/2$. The rest of the fields, $A, B$ are expected to have the same scaling dimension due to the
$SU(2)$ symmetry that acts on the index $i$ of both $A_i$ and $B_i$, as well as the $Z_2$
symmetry of the quiver theory that exchanges $A$ and $B$ together with charge conjugation. Since
the superpotential is of the form $\phi A B$ and has a scaling 2, we conclude that $A$ and $B$
have a scaling dimension $3/4$.

Let us confirm this with a more detailed analysis. The permanent of the Kasteleyn matrix shows
that there are 5 perfect matchings $p_\alpha, \alpha=1\ldots5$.  From the point of view of the
2d toric diagram, $p_1, p_2, p_3$ correspond to the external points on ${\cal T}_2$ in
\fref{cctd3d} while $p_4, p_5$ corresponds to the single point on the boundary of  ${\cal T}_2$
of \fref{cctd3d}, in between $p_1$ and $p_2$. In the 3d toric diagram ${\cal T}_3$, $p_4,p_5$
split.

On the coherent component of the moduli space
we find the parameterization of fields in terms of perfect matchings  \cite{Franco:2005rj}
\beq
\label{c2pms} A_1 = p_1 p_4, \quad A_2 = p_2 p_4, \quad B_1 = p_1 p_5, \quad B_2 = p_2 p_5,
\quad\phi_1 = \phi_2 = p_3 .
\eeq
The previous relation is easily obtained from the permanent of the Kasteleyn matrix with the
insertion of elementary fields
\be
 {\rm perm}     \left(
\begin{array}{cc}
 A_1 z^{-1}+ B_2  z x^{-1} & \phi_2  \\
 \phi_1 y &B_1 z+  A_2 z^{-1} x^{-1}
\end{array}
\right)
 = A_1 B_1+ A_1 A_2  \frac{1}{z^2 x} + B_1 B_2 \frac{z^2}{x} + B_2 A_2 \frac{1}{x^2} +\phi_1 \phi_2  y .
\label{pmc2z2}
\ee
by recalling that each monomial corresponds to a perfect matching. For example $A_1$ belongs to
the perfect matchings $p_1$ and $p_4$ and this leads to the relation $A_1=p_1 p_4$ and similarly
for the other fields.

From this we learn that the external points of ${\cal T}_2$, denoted $p_1, p_2, p_3$, carry the
information on the complex directions of $\IC^2/\IZ_2\times \IC$, while the internal points of
${\cal T}_2$, denoted $p_4, p_5$ are carrying the information on the $A$ quantum number and $B$
quantum number, respectively, namely, $p_4$ counts the number of $A$'s and $p_5$ counts the
number of $B$'s. All together there are 4 conserved charges corresponding to each point in the
toric diagram ${\cal T}_2$. The two perfect matchings $p_4, p_5$ are associated with the
internal point and therefore do not carry a quantum number which is independent. Instead we can
introduce a conserved quantum number which in 3+1 dimensions is called the baryonic charge which
counts the number of $A$'s minus the number of $B$'s. In 2+1 dimensions all 4 charges are
mesonic. When the number of gauge groups is more than 2 we can find additional baryonic charges
and they play a role in the next set of examples.

The moduli space is given by the collection of these 5 perfect matchings subject to the relation
$p_1+p_2=p_4+p_5$ which give rise to the charge vector $(1,1,0,-1,-1)$. We see that the quiver
fields $\phi$'s, $A$'s and $B$'s are gauge invariants with respect to this charge. Furthermore,
\eref{c2pms} gives the conifold relation
\beq
\label{c2rel}
A_1 B_2 = A_2 B_1.
\eeq
This relation is the F-term relation corresponding to the adjoint fields $\phi$. To compute the
fully refined Hilbert series it is enough for this case to use the property that the moduli
space ${\cal C}\times\IC$ \cite{Hanany:2008cd} is a complete intersection. We further exploit
the global symmetry of this model, which coincides with the symmetry of the moduli space,
$SU(2)_1\times SU(2)_2\times U(1)_1\times U(1)_2$. Let us introduce fugacities $t_{1,2}$ for
$U(1)_{1,2}$, and $x_{1,2}$ for $SU(2)_{1,2}$, respectively. With these fugacities and with the
help of the charge vector we assign fugacities $t_1 x_1, t_1/x_1, t_2, t_1 x_2, t_1/x_2$ to the
perfect matchings, $p_1, p_2, p_3, p_4, p_5$, respectively. From this and from \eref{c2pms} it
follows that the quiver fields $\phi_1, A_1, B_1, \phi_2, A_2, B_2$ have fugacities $t_2, t_1^2
x_1 x_2, t_1^2 x_2/x_1, t_2, t_1^2x_1/x_2, t_1^2/x_1 x_2$, respectively. This assignment puts
$p_{1,2}$ and $p_{4,5}$ in the $[1;0]$ and $[0,1]$ representations of $SU(2)_1\times SU(2)_2$,
respectively, while the A's and B's transform in the $[1;1]$ representation. With these
preparations we are now ready to write down the refined Hilbert series and it takes the form
\beq
\label{c2z2ref} g \left (t_1,t_2, x_1, x_2; \widetilde{\IC^2/\IZ_2\times \IC}_{\{1,-1\}} \right
) = \frac{1-t_1^4}{(1 - t_1^2 x_1 x_2)(1 - t_1^2 x_2 / x_1)(1-t_2)(1-t_1^2 x_1 / x_2)(1 - t_1^2 / x_1 x_2)} ,
\eeq
where the denominator takes into account the generators, $A$'s, $B$'s and $\phi$, and the
numerator takes into account the relation \eref{c2rel}. Alternatively, for generalization to
cases in which the moduli space is not a complete intersection we can use a Molien integral,
\beq\label{mol}
g \left (t_1,t_2, x_1, x_2; \widetilde{\IC^2/\IZ_2\times \IC}_{\{1,-1\}} \right ) =
\oint_{|z|<1} \frac{d z}{2\pi i z} \frac{1}{(1-t_1 x_1 z)(1-t_1 z/x_1)(1-t_2)(1-t_1
x_2/z)(1-t_1/x_2 z)} .
\eeq
In this formula the denominator takes into account the six perfect matchings and
the integration takes into account the existence of the linear relation
$p_1+p_2=p_4+p_5$.

Next we are ready to compute the volume of the SE$_7$. Set $t_1 =e^{-\mu b_1}, t_2=e^{-\mu
b_2}$. Due to the non-abelian symmetry $SU(2)_1\times SU(2)_2$ we do not expect the
corresponding fugacities to affect the formula for the volume and can therefore safely set their
values to 1 at the extremal point. Alternatively, if this argument is not trusted, we can assume
a dependence on these two variables, get a more complicated formula, and find that the extremum
at $x_1=x_2=1$ follows. Taking the limit we find
\beq
\lim_{\mu\rightarrow0} \mu^4 g \left (e^{-\mu b_1}, e^{-\mu b_2}, 1, 1;
\widetilde{\IC^2/\IZ_2\times \IC}_{\{1,-1\}} \right ) = \frac{1}{ 4 b_1^3 b_2}
\eeq
The CY condition now sets the sum over all external perfect matchings in the 3d toric diagram
${\cal T}_3$ to be $4b_1+b_2=2$, from which the minimization gives $b_1=3/8, b_2 = 1/2$. The
four points $p_1,p_2,p_4,p_5$ have charge $3/8$, equal as expected by symmetry of the toric
diagram, and $p_3$ has charge $1/2$. Tracing this back to the quiver fields we find a reassuring
result that the $R$ charge for the adjoint fields $\phi_{1,2}$ is $1/2$ consistent with the
canonical dimension for a scalar field in 2+1 dimensions. The fields $A$ and $B$ have charge
$3/4$ which is consistent with the cubic superpotential.

Since there are no baryonic charges, the R-charges of the fields are uniquely determined by the
minimization on the mesonic Hilbert series. For consistency we check that the values of charges
for $p_i$ are given by the normalized volumes of the corresponding divisors $D_i$ in the
geometry. By symmetry, $D_1,D_2,D_4,D_5$ will have the same volume. The normalized volumes can
be computed as explained in the Appendix.  The explicit computation is actually superfluous
because we expect result $1/2$ for the free factor $\mathbb{C}$ corresponding to $D_3$; the fact
that normalized volumes add up to $2$ then fix the value of the others to $3/8$. Let us do
nevertheless the explicit computation to prepare for more complicated examples. Consider the
Molien integral (\ref{mol}). The five terms in the denominator are the weights corresponding to
the five perfect matchings $p_i$ and therefore to the five divisors $D_i$. Formula
(\ref{divisors}) instructs us to compute the Hilbert series corresponding to $D_i$ by computing
the same Molien integral with the insertion of (the inverse of) the weight corresponding to the
divisor $D_i$. We have (setting by symmetry $x_1=x_2=1$),
\bea
g \left (D_{1,2}\, ; t_1,t_2; \widetilde{\IC^2/\IZ_2\times \IC}_{\{1,-1\}} \right ) &=&
\oint_{|z|<1} \frac{d z}{2\pi i z} \frac{(t_1 z)^{-1}}{(1-t_1 z)^2(1-t_2)(1-t_1/z)^2 }, \nonumber \\
g \left (D_{4,5}\, ; t_1,t_2; \widetilde{\IC^2/\IZ_2\times \IC}_{\{1,-1\}} \right ) &=&
\oint_{|z|<1} \frac{d z}{2\pi i z} \frac{(t_1 /z)^{-1}}{(1-t_1 z)^2(1-t_2)(1-t_1/z)^2 }, \nonumber \\
g \left (D_{3}\, ; t_1,t_2; \widetilde{\IC^2/\IZ_2\times \IC}_{\{1,-1\}} \right ) &=&
\oint_{|z|<1} \frac{d z}{2\pi i z} \frac{t_2^{-1}}{(1-t_1 z)^2(1-t_2)(1-t_1/z)^2 } .
\nonumber \\
\eea
The normalized volumes are then
\bea
\frac{g \left (D_{i}\, ; e^{-\mu b_1}, e^{-\mu b_2}, 1, 1; \widetilde{\IC^2/\IZ_2\times
\IC}_{\{1,-1\}} \right )}{g  \left (e^{-\mu b_1}, e^{-\mu b_2}, 1, 1;
\widetilde{\IC^2/\IZ_2\times \IC}_{\{1,-1\}} \right )}&=& 1 + \frac{3}{8} \mu +\cdot\cdot\cdot
\qquad i=1,2,4,5 ,
\nonumber \\
\frac{g \left (D_{3}\, ; e^{-\mu b_1}, e^{-\mu b_2}, 1, 1; \widetilde{\IC^2/\IZ_2\times
\IC}_{\{1,-1\}} \right )}{g  \left (e^{-\mu b_1}, e^{-\mu b_2}, 1, 1;
\widetilde{\IC^2/\IZ_2\times \IC}_{\{1,-1\}} \right )}&=& 1 + \frac{1}{2} \mu +\cdot\cdot\cdot
\eea
as expected. From (\ref{c2pms}) we recover the previous assignment of R-charges for the
elementary fields. Notice that no elementary field carries the charge $3/8$ corresponding to the
volumes of the four divisors $D_1,D_2,D_4,D_5$. $\phi$ is associated with $D_3$. The elementary
fields $A$ and $B$ seem to be associated, consistently with the tiling prescription
\eref{c2pms}, to combinations of divisors $D_1+D_4, D_2+D_4,D_1+D_5,D_2+D_5$. It remains to
explain whether there are BPS states in the theory associated to M5 branes wrapped on single
divisors.

In view of the numerical values for the R-charges we can now define two new fugacities: $t$
which counts the total R-charge, and $q$ which is a conserved current that satisfies the
condition that the total charge over external points of the 3d toric diagram is 0. We find
$t_1^2 = t^3 q, t_2 = t^4/q^2,$ such that the new form of the partition function \eref{c2z2ref}
is
\beq
g \left (t,q, x_1, x_2; \widetilde{\IC^2/\IZ_2\times \IC}_{\{1,-1\}} \right ) = \frac{1-t^{6}
q^2}{(1-t^3 q x_1 x_2)(1-t^3 q x_2 / x_1)(1-t^4 / q^2)(1-t^3 q x_1 / x_2)(1-t^3 q / x_1 x_2)} .
\eeq
We summarize the collection of charges in Table \ref{c2z2cha}.
\begin{table}[htdp]
\caption{Global charges for the perfect matchings for the quiver gauge theory on
the M2 brane probing the modified $ \widetilde{\IC^2/\IZ_2\times \IC} $ singularity.}
\begin{center}
\begin{tabular}{|c||c|c|c|c|c|c||c|}
\hline
 & $SU(2)_1$    & $SU(2)_2$ & $U(1)_q$ & $U(1)_R$ & fugacities \\ \hline \hline
$p_1$ & $1$     & $0$   & $1/2$ & $3/8$ & $t_1 x_1$ \\ \hline $p_2$ & $-1$    & $0$   & $1/2$ &
$3/8$ & $t_1 / x_1$ \\ \hline $p_3, \phi_1, \phi_2$ & $0$     & $0$   & $-2$ & $1/2$ & $t^4 / q^4$ \\ \hline
$p_4$ & $0$     & $1$   & $1/2$ & $3/8$ & $t_1 x_2$ \\ \hline $p_5$ & $0$     & $-1$  & $1/2$ &
$3/8$ & $t_1 / x_2$ \\ \hline
$A_1$ & $1$     & $1$  & $1$ & $3/4$ & $t^3 q x_1 x_2$ \\ \hline
$B_1$ & $1$     & $-1$  & $1$ & $3/4$ & $t^3 q x_1 / x_2$ \\ \hline
\end{tabular}
\end{center}
\label{c2z2cha}
\end{table}

For higher values of the CS couplings, $k, -k$ the moduli space can be determined as follows. It
is a $\IZ_k$ action on ${\cal C}\times \IC$ and the $\IZ_k$ action is easiest to see on the
quiver fields, $\phi, A, B$, with charges $0, 1, -1$, respectively. Using Table \ref{c2z2cha}
this identifies with the weight of $SU(2)_2$, implying that $x_2$ is acted by $\IZ_k$ and all
other weights are free. The resulting Hilbert series is
\bea
& & g \left (t,q, x_1, x_2; \widetilde{\IC^2/\IZ_2\times \IC}_{\{k,-k\}} \right ) =
\\ \nonumber &=& \frac{1-t^6 q^2}{k} \sum_{j=0}^{k-1} \frac{1}{(1 - \omega^j t^3 q x_1 x_2)(1 - \omega^j t^3 q x_2 / x_1)(1-t^4 / q^2)(1- \omega^{-j} t^3 q x_1 / x_2)(1- \omega^{-j} t^3 q / x_1 x_2)} .
\eea

\subsection{The $\widetilde{L^{aba}}$ theories}

We proceed with the analysis of more models of the type $L^{aba}$
\cite{Cvetic:2005ft,Cvetic:2005vk,Martelli:2005wy,Benvenuti:2005ja,Franco:2005sm,Butti:2005sw}.
A special subset of these theories are those with $a=0$ corresponding to orbifolds of $\IC^3$
with higher amount of supersymmetry in 3+1 dimensions, namely with 8 supercharges. Note,
however, that while this is true in 3+1 dimensions, the presence of the CS couplings breaks the
supersymmetry down to 4 supercharges in 2+1 dimensions, as can readily be seen by the simplest
family of this class corresponding to $b=2$ of the previous subsection.

The method used in the analysis of the previous subsection is with the explicit form of perfect
matchings and their relation to the basic fields of the quiver. This method does not seem to be
easily generalized to higher orbifold cases since the number of perfect matching grows
exponentially with the order of the singularity $b$ and the treatment using these degrees of
freedom becomes cumbersome. Luckily we have a simple property of these models which saves the
day. The coherent component of the master space for all of this class of theories is a complete
intersection which is generated by $2a+2b$ variables that are subject to $a+b-2$ constraints. We
further have luck on our side and use the large hidden symmetry that the master space has. The
hidden symmetry of the coherent component of the master space is $SU(a)\times SU(b)$ which
together with the explicit symmetry directions that add $U(1)^4$, one of which is baryonic and
the rest are mesonic, leads to a global symmetry of rank $a+b+2$ as expected from the dimension
of the coherent component of the moduli space. We introduce 4 chemical potentials $t_1, t_2,
t_3, t_4$ for the $U(1)$'s and chemical potentials $x_i, y_j$ for the hidden symmetries. In the
case that $a=0$ there are only 3 $U(1)$'s, corresponding to 3 mesonic charges and the hidden
symmetry group is $SU(b)$. Together they form a global symmetry of rank $b+2$ which is again the
expected value. The $2a+2b$ generators transform as $[1,0,\ldots, 0; 0, \dots, 0]$, $[0,\ldots,
0,1; 0, \dots, 0]$, $[0,\ldots, 0; 1, 0, \dots, 0]$, $[0,\ldots, 0; 0, \dots, 0, 1]$ of
$SU(a)\times SU(b)$. Each of these representations carries one of the 4 $U(1)$ charges. The
relations are singlets of the hidden symmetries and carry charges with respect to two of the
U(1)'s. With this information it is now possible to write the refined Hilbert series for the
coherent component of the master space.
\bea
g(t_i, & x_j &, y_k; L_{aba}) = (1-t_1 t_2)^{a-1}(1-t_3 t_4)^{b-1} \times \\
& PE & \left [ t_1 [1,0,\ldots, 0; 0, \dots, 0] + t_2 [0,\ldots, 0,1; 0, \dots, 0] + t_3 [0,\ldots, 0; 1, 0, \dots, 0] + t_4 [0,\ldots, 0; 0, \dots, 0, 1] \right ]\nonumber
\eea
For the case of $a=0, b=n$ this equation is slightly modified and takes the form
\beq
g(t_i, x_j; \IC^2/\IZ_n \times \IC) =\frac{(1-t_1 t_2)^{n-1}}{(1-t_3)} PE \left [ t_1 [1,0,\ldots, 0] + t_2 [0,\ldots, 0,1] \right ]
\eeq

\subsection{The $\widetilde{\IC^2/\IZ_3\times \IC}$ theories}\label{c2z3sec}

\begin{figure}[ht]
\begin{center}
  \includegraphics[totalheight=1.5cm]{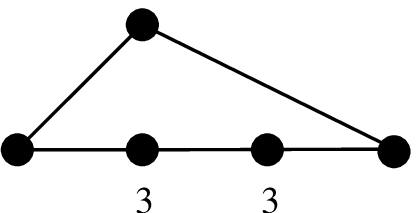}
  \caption{The 2d toric diagram for $\IC_2 / \IZ_3 \times \IC$.}
  \label{c2z3_2d}
\end{center}
\end{figure}

Let us specialize to the case of $a=0, b=3$. The superpotential of the theory is
\beq W = \phi_1 (A_1 B_1 -B_3 A_3) +\phi_2 (A_2 B_2 -B_1 A_1) +\phi_3 (A_3 B_3 -B_2 A_2) \, .\eeq

The Hilbert series for the coherent component of the master space is
\beq
g(t_1,t_2,t_3, x_1,x_2; \IC^2/\IZ_3 \times \IC) =\frac{(1-t_1 t_2)^{2}}{(1-t_3)} PE \left [ t_1 [1,0] + t_2 [0,1] \right ] .
\eeq
Since this is a complete intersection moduli space it is possible to write the relations explicitly,
\beq
\label{c2z3rel}
A_1 B_1 = A_2 B_2 = A_3 B_3.
\eeq
Let us fix the CS coefficients to be $k_1, k_2, -k_1-k_2$. Before treating the general case let
us first take $k_1=k, k_2=0.$ To get the moduli space of the 2+1 dimensional theory we need to
divide by the D term of the gauge group with CS equal to 0. We need to identify the charges of
the 6 generators under this gauge group. This is done in Table \ref{c2z3char}.
\begin{table}[htdp]
\caption{Gauge charges for the generators of the theory on
an M2 brane probing the modified $ \widetilde{\IC^2/\IZ_3\times \IC} $ singularity.}
\begin{center}
\begin{tabular}{|c||c|c|c|c|c|c||c||c|}
\hline
 & $U(1)_1$    & $U(1)_2$ & $U(1)_3$ & $U(1)_R$ & fugacities & p.m.\\ \hline \hline
$A_1$ & $1$    & $-1$   & $0$ & $3/4$ & $t_1 x_1$ & $p_1 q_1 \tilde q_2\tilde q_3 $\\ \hline
$A_2$ & $0$    & $1$   & $-1$ & $3/4$ & $t_1 x_2 / x_1$ & $p_1 \tilde q_1 q_2 \tilde q_3 $ \\ \hline
$A_3$ & $-1$   & $0$   & $1$ & $3/4$ & $t_1 / x_2$ & $ p_1 \tilde q_1 \tilde q_2 q_3 $ \\ \hline
$B_1$ & $-1$   & $1$   & $0$ & $3/4$ & $t_2 / x_1$ & $p_2 \tilde q_1  q_2 q_3 $ \\ \hline
$B_2$ & $0$    & $-1$  & $1$ & $3/4$ & $t_2 x_1 / x_2$ & $ p_2 q_1 \tilde q_2 q_3$ \\ \hline
$B_3$ & $1$    & $0$  & $-1$ & $3/4$ & $t_2 x_2$ & $p_2 q_1 q_2 \tilde q_3$\\ \hline
$\phi_1, \phi_2, \phi_3$ & $0$ & $0$&$0$& $1/2$& $t_3$ & $p_3$\\ \hline
\end{tabular}
\end{center}
\label{c2z3char}
\end{table}
A quick inspection shows that we need to integrate over the $x_1$ variable. Let us write the
Hilbert series for the coherent component of the master space more explicitly, together with the
integral
\bea
\nonumber
&&g \left (t_1,t_2,t_3, x_2; \widetilde{\IC^2/\IZ_3 \times \IC}_{\{1,0,-1\}} \right ) = \\ \nonumber
&=& \oint_{|x_1|<1}\frac{dx_1}{x_1}\frac{(1-t_1 t_2)^{2}}{(1-t_3)(1-t_1 x_1)(1-t_1 x_2/x_1)(1-t_1/x_2)(1-t_2/x_1)(1-t_2x_1/x_2)(1-t_2 x_2)} \\
&=& \frac{1-t_1^2 t_2^2}{(1-t_1^2 x_2)(1-t_1/x_2)(1-t_2^2/x_2)(1-t_2 x_2)(1-t_3)} .
\eea
The mesonic branch of the 2+1 dimensional theory is then a complete intersection moduli space
generated by 5 generators which are subject to one relation. Explicitly, we can find the
generators to be $\phi_1=\phi_2=\phi_3$ parametrizing $\IC$ and $A_3, B_3, M_1 = A_1 A_2, M_2 =
B_1 B_2.$ Using \eref{c2z3rel} the relation can also be explicitly written,
\beq
M_1 M_2 = A_3^2 B_3^2.
\eeq
This identifies the moduli space as the $Z_2$ orbifold of the conifold, otherwise known as
$L^{222}$. To summarize, the mesonic moduli space of $\widetilde{\IC^2/\IZ_3 \times
\IC}_{\{1,0,-1\}}$ is $L^{222}\times\IC$.

\begin{figure}[ht]
\begin{center}
  \includegraphics[totalheight=3.5cm]{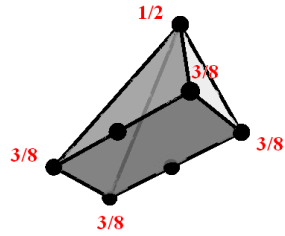}
  \caption{The 3d toric diagram for $L^{222}\times\IC$. The R-charges are shown in red.}
  \label{td3dl222}
\end{center}
\end{figure}

We are now ready to compute the R-charges for the fields. On the mesonic branch of the 2+1
dimensional theory we just need to compute and minimize the volume of the seven manifold. This
is done as usual by setting $t_1=e^{- \mu b_1},t_2 = e^{-\mu b_2},t_3=e^{-\mu (2-b_1-b_2)},
x_2=e^{-\mu b_x}$ and computing the coefficient of the pole $1/\mu^4$. The restriction on the
exponents comes from the fact that the holomorphic top form scales as $t_1  t_2 t_3$. We obtain
the volume function
\beq
Z(b_1,b_2,b_x)=\frac{2(b_1+b_2)}{(b_1+b_2-2)(2 b_2^2 +b_2 b_x-b_x^2)(-2 b_1^2 +b_1 b_x +b_x^2)}
\eeq
whose minimization gives $b_x=0,b_1=b_2=3/4$ corresponding to a dimension $3/4$ for the fields
$A$ and $B$ and $1/2$ for $\phi$. These are recorded in Table \ref{c2z3char}. As in subsection
\ref{c2z2sec} the charges can be computed by just using symmetry arguments and simple scalings.
For reference, we notice that the toric diagram of $L^{222}\times\IC$ has five external points
corresponding to five volumes; as usual the factor $\IC$ has volume $1/2$ and by symmetry and
the fact that the normalized volumes add up to $2$ we obtain the value $3/8$ for the other four.
With the parameterization given in Table \ref{c2z3char} the five points are labeled by $p_1,
p_2,p_3,q_3,\tilde q_3$ and these values reproduce the R-charges of the elementary fields.

For generic CS parameters we need to divide by the gauge group $k_2 U(1)_1-k_1 U(1)_2$. The
refined Hilbert series takes the form
\bea
&&g \left (t_1,t_2,t_3, x; \widetilde{\IC^2/\IZ_3 \times \IC}_{\{k_1,k_2,-k_1-k_2\}} \right ) = \\ \nonumber
&=& \oint_{|w|<1}\frac{dw}{w}\frac{(1-t_1 t_2)^{2}}{(1-t_3)(1-t_1 w^{k_1+k_2})(1-t_1 x/w^{k_1})(1-t_1/x w^{k_2})(1-t_2/w^{k_1+k_2})(1-t_2 w^{k_1}/x)(1-t_2 w^{k_2} x)}
\eea
which corresponds to the change of variables $x_1 = w^{k_1+k_2}, x_2 = x w^{k_2}$ from Table
\ref{c2z3char} and a subsequent integration over $w$ which is the weight under the gauge group
$k_2 U(1)_1-k_1 U(1)_2$. We further need the fugacities for the perfect matchings. Using Table
\ref{c2z3char} we can choose
\beq
p_1 : t_1, \,\,\, p_2 : t_2, \,\,\, p_3 : t_3, \,\,\,
q_1: \frac{x_1}{x_2},\,\,\,
q_2: \frac{x_2}{x_1},\,\,\,
q_3 : 1,\, \,\,
\tilde q_1 : \frac{1}{x_2},\,\,\,
\tilde q_2 : 1,\,\,\,
\tilde q_3 : x_2
\eeq

An interesting case is $k_1=k_2=1$ which has $SU(2)$ global symmetry. $x$ is the corresponding
weight. The 3d toric diagram has a hexagonal base and seven external points, which can set in
correspondence with $p_1,p_2,p_3,q_1,q_3,\tilde q_1,\tilde q_3$. $q_2$ and $\tilde q_2$ are
still internal point in the 3d toric diagram. A not difficult but long computation with the
formulae in the Appendix give the following values for the volumes: $1/2$ for $p_1$, $3/10$ for
$p_2,p_3$ and $9/40$ for $q_1,q_3,\tilde q_1,\tilde q_3$. As seen from the table this still
correspond to R-charge $3/4$ for $A,B,C$ and R-charge $1/2$ for $\phi$.

\begin{figure}[ht]
\begin{center}
  \includegraphics[totalheight=3.0cm]{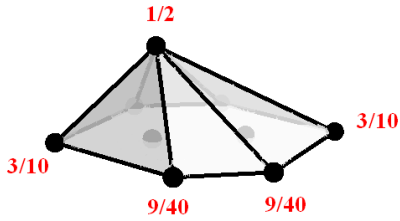}
  \caption{The 3d toric diagram for $\widetilde{\IC^2/\IZ_3 \times \IC}_{\{1,1,-2\}}$. The R-charges are shown in red.}
  \label{td3d112}
\end{center}
\end{figure}

\subsubsection{A family of $\widetilde{\IC^2/\IZ_{n+1}\times \IC}$ theories}

An interesting family of theories can be taken to be $\widetilde{\IC^2/\IZ_{n+1}\times
\IC}_{\{1,0,\ldots,0,-1\}}$. The first member of this class is studied in Section \ref{c2z2sec}
corresponding to $\widetilde{\IC^2/\IZ_{2}\times \IC}_{\{1,-1\}}$ with a 2+1 dimensional moduli
space $L^{111}\times \IC$, where we have written the conifold ${\cal C} = L^{111}$ in a form
which is more suitable for the generalization to higher $n$. The second member of this class is
discussed in Section \ref{c2z3sec} for $n=2$. The general $n$ case gives a moduli space which is
$L^{nnn}\times \IC$. This can be summarized with the following points. There are $3n+3$ fields
$\phi_i, A_i, B_i, \quad i=1\ldots n+1$. The superpotential is
\beq
W = \sum_{i=1}^{n+1} \phi_i (A_i B_i - B_{i-1} A_{i-1}) ,
\eeq
where the index $i$ is taken to be cyclic, modulo $n+1$. The F-term equations on the coherent
component of the master space take the form
\beq
A_1 B_1 = A_2 B_2 = \cdots = A_n B_n = A_{n+1} B_{n+1},
\eeq
giving a complete intersection moduli space generated by $2n+2$ variables subject to $n$
constraints and another copy of $\IC$, generated by $\phi_1=\cdots=\phi_{n+1}$. A choice of CS
coefficients, ${\{1,0,\ldots,0,-1\}}$ instructs to mod out by the $n-2$ gauge groups which have
CS level equal to $0$. The resulting Hilbert series for the CY four-fold is
\bea
\nonumber
g \left (t_1,t_2,t_3, x; \widetilde{\IC^2/\IZ_{n+1} \times \IC}_{\{1,0,\ldots,0,-1\}} \right ) =
 \frac{1-t_1^n t_2^n}{(1-t_1^n x)(1-t_1/x)(1-t_2^n/x)(1-t_2 x)(1-t_3)} .
\eea
The gauge invariant generators of this moduli space are
\beq
M_1 = \prod_{i=1}^n A_i \quad, M_2 = \prod_{i=1}^n B_i, \quad A_{n+1}, \quad B_{n+1}, \quad \phi,
\eeq
and the relation they satisfy is
\beq
M_1 M_2 = A_{n+1}^n B_{n+1}^n
\eeq
corresponding to the announced result that the moduli space is $L^{nnn}\times \IC$, where
$L^{nnn}$ is a non-chiral $\IZ_n$ orbifold of the conifold. The scaling dimensions can be
determined by symmetry. There are 5 external points in the toric diagram of $L^{nnn}\times \IC$,
one corresponds to $\phi$ which has a scaling $1/2$. The other 4 are completely symmetric and
get scaling dimension $3/8$ each. Finally, each of the fields $A, B$ have a scaling dimension
$3/4$.
\subsection{The $\widetilde{SPP}$ ($L^{121}$) revisited}


\begin{figure}[ht]
\begin{center}
  \hskip -6cm
  \includegraphics[totalheight=3cm]{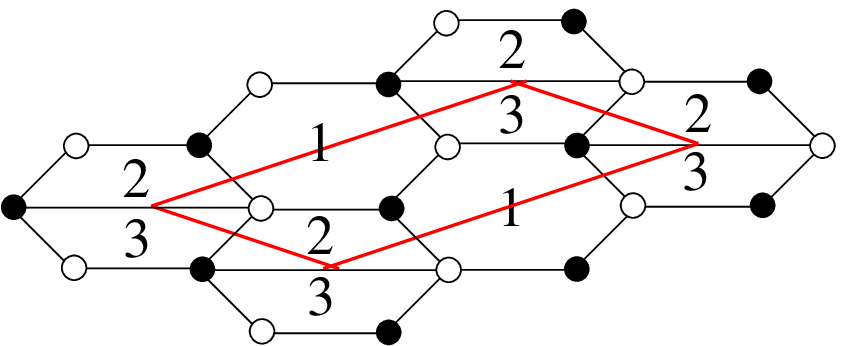}
  \vskip -3cm
  \hskip 9cm
  \includegraphics[totalheight=3cm]{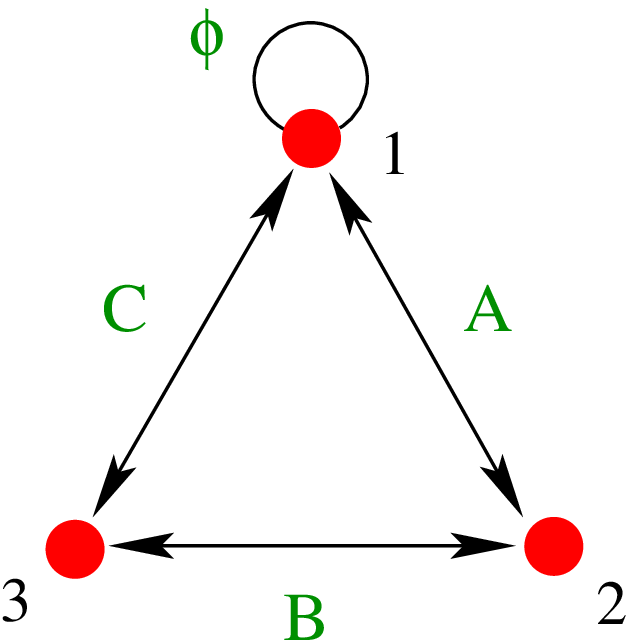}
  \caption{(i) Brane tiling for $\widetilde{SPP}$. The fundamental domain is shown in red. 
     \ (ii) The corresponding quiver.}
  \label{spp1}
\end{center}
\end{figure}

\begin{figure}[ht]
\begin{center}
  \includegraphics[totalheight=1.5cm]{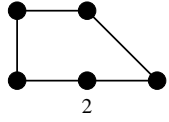}
  \caption{Toric diagram of the Suspended Pinch Point singularity ($xy=uv^2$). }
  \label{spp_toric}
\end{center}
\end{figure}

The tiling and toric diagram are given in \fref{spp1}. The 2d toric diagram is given in
\fref{spp_toric}. The theory has  chiral fields $\phi,A_i,B_i,C_i$ indicated in \fref{spp1} and
interacting with the superpotential
\beq W=  \phi ( A_1 A_2  - C_2 C_1) - A_2 A_1 B_1 B_2 + C_1 C_2 B_2 B_1 . \eeq

To compute the 3d toric diagram we write the Kasteleyn matrix  (with all the fields inserted)
\be
K =   \left(
\begin{array}{cc}
C_1 z^{{k_1}-{k_3}}+ C_2 z^{-{k_1}+{k_3}} y &\ B_1  x^{-1} z^{-{k_2}+{k_3}}+ B_2 x^{-1} y  z^{{k_2}-{k_3}}  \\
 \phi                              & \  A_1 z^{-{k_1}+{k_2}}+ A_2 z^{{k_1}-{k_2}} y
\end{array}
\right) ,
\ee
and the permanent gives
\bea
  \textrm{perm} \, K =  A_1 C_1 z^{{k_2}-{k_3}}+ B_1 \phi \frac{z^{-{k_2}+{k_3}}}{x}+ C_1 A_2 z^{2 {k_1}-{k_2}-{k_3}}
  y+ A_1 C_2 z^{-2 {k_1}+{k_2}+{k_3}} y \nonumber \\
  + B_2 \phi \frac{z^{{k_2}-{k_3}} y}{x}+ A_2 C_2 z^{-{k_2}+{k_3}} y^2 .
\eea

Let us denote the six perfect matchings corresponding to the six monomials in the previous
expression as $p_1,p_3,q_2,q_1,p_4,p_2$ in the given order. From this expression we see that,
for example, $A_1$ belongs to the perfect matchings $p_1$ and $q_1$ and therefore it can be
parametrized as $p_1 q_1$. A similar computation for the other fields gives the result in Table
\ref{spptab}.

\begin{table}[h]
\caption{Fields, perfect matchings and fugacities for the quiver gauge theory on
the M2 brane probing the modified $ \widetilde{SPP}$ singularity.}
\begin{center}
\begin{tabular}{|c||c|c|}
\hline
& p.m & fugacities \\ \hline\hline
$A_1$ & $p_1 q_1$ &$ t_1^2 x_1 x_3$ \\ \hline
$A_2$ & $p_2 q_2$ & $t_1^2/ x_1 x_3$\\ \hline
$C_1$ & $p_1 q_2$ & $t_1^2 x_1/ x_3$\\ \hline
$C_2$ & $p_2 q_1$ & $t_1^2 x_3/ x_1$\\ \hline
$B_1$ & $p_3$ & $t_2 x_2$\\ \hline
$B_2$ & $p_4$ & $t_2/x_2$\\ \hline
$\phi$ & $p_3 p_4$  & $t_2^2$\\ \hline
\end{tabular}
\end{center}
\label{spptab}
\end{table}
The coherent component of the master space is the space of the 6 perfect matchings which are
subject to the single relation $p_1+p_2=q_1+q_2$. This variety is $\mathbb{C}^6$, parametrized
by $p_1,p_2,p_3,p_4,q_1,q_2$ divided by the action $(1,1,0,0,-1,-1)$ and it coincides with the
conifold times  $\mathbb{C}^2$. This can be easily seen also from the F-term equations
\beq A_1 A_2 =C_1 C_2\, , \qquad\qquad \phi= B_1B_2  \eeq
The symmetry of this space is $SU(2)_1\times SU(2)_2 \times SU(2)_3 \times U(1)_1 \times
U(1)_2$, where all three $SU(2)$'s are hidden \cite{Forcella:2008bb}. We introduce five
independent weights $t_i,x_i$ corresponding to the global symmetries of the master space as
recorded in the table ($t$'s for $U(1)$'s and $x$'s for $SU(2)$'s). The perfect matchings
$p_1,p_2,p_3,p_4,q_1,q_2$ have weights $t_1 x_1,t_1/x_1,t_2 x_2,t_2/x_2,t_1 x_3,t_1/x_3$.
  The Hilbert series for the master space is then
\bea
g(t_i,x_i; SPP) &=&\int \frac{dz}{2\pi i z} \frac{1}{(1-t_1 z x_1)(1-t_1 z / x_1)(1-t_2 x_2)(1-t_2/x_2)(1-t_1 x_3 / z)(1-t_1/ x_3 z)} \nonumber\\
&=& \frac{1-t_1^4}{(1-t_1^2 x_1 x_3) (1-t_1^2 x_1/ x_3) (1-t_1^2 x_3/ x_1) (1-t_1^2/ x_1 x_3) (1-t_2 x_2)(1-t_2/x_2)}
\label{hsSPP}
\eea
which is indeed the Hilbert series for the conifold times $\mathbb{C}^2$.

The Newton polygon of $ \textrm{perm} \, K |_{z=1}$ gives the Suspended Pinch Point toric
diagram (\fref{spp_toric}).
By turning on the ``magnetic flux'', the points of the diagram get pulled into the third
dimension. Depending on the choices for $k_i$, we have several possibilities:

\subsubsection {(i) $(k_1, k_2, k_3) = (1,-1, 0)$} 
\begin{figure}[ht]
\begin{center}
  \hskip -7cm
  \includegraphics[totalheight=3.0cm]{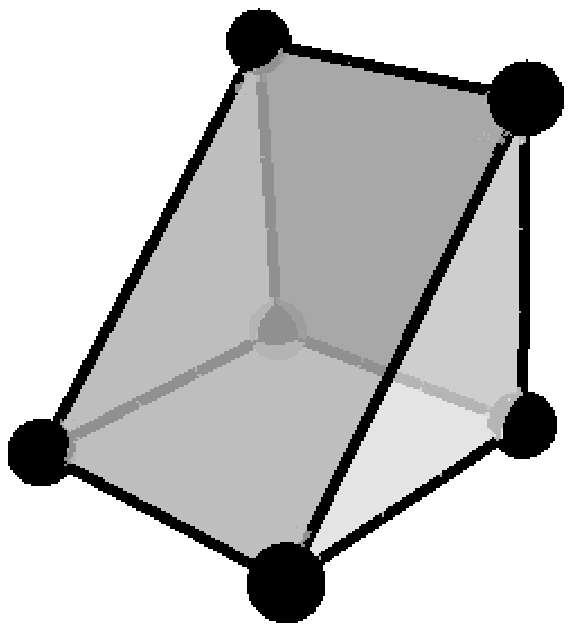}
  \vskip -3.3cm
  \hskip 7cm
  \includegraphics[totalheight=3.5cm]{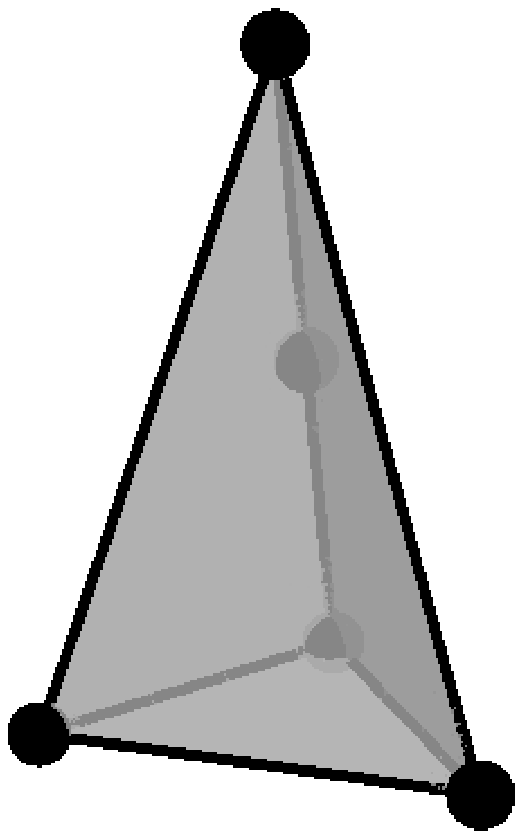}
  \caption{(i) Toric diagram for $D_3$. \ (iii) Toric diagram for $\IC^2 / \IZ_2 \times  \IC^2$.}
  \label{td3d_3}
\end{center}
\end{figure}

gives the $D_3$ model whose
 3d toric diagram is depicted in \fref{td3d_3}(i): the two points $q_i$ split
and we obtain a toric diagram with six external points.

We need to mod out by the third gauge group. The Hilbert series for $D_3$ takes the form
\beq
g \left (t_1,t_2, x, y; \widetilde{SPP}_{\{1,-1,0\}} \right ) =\frac{1-t_1^4 t_2^2}{(1-t_1^2 x)(1-t_1^2/x)(1-t_1^2 t_2 y)(1-t_1^2 t_2/y)(1-t_2^2)} ,
\eeq
where $x = x_1 x_3$ and $y = x_1 x_2 / x_3$ are two new $SU(2)_x\times SU(2)_y$ weights, this
group being a particular subgroup of the hidden $SU(2)_1\times SU(2)_2\times SU(2)_3$ of the
master space of SPP. This moduli space is a complete intersection of dimension 4 generated by
the 5 $U(1)_3$ invariants
\beq
A_1, \quad A_2, \quad M_1 = B_1 C_1, \quad M_2 = C_2 B_2, \quad M_3 = B_1 B_2,
\eeq
which satisfy the relation
\beq
M_1 M_2 = A_1 A_2 M_3.
\eeq
The mesonic spectrum can be computed by taking the scaling $t_1 = e^{-\mu b_1}, t_2 = e^{-\mu
b_2}, x=1, y=1$. The original top holomorphic form on $\mathbb{C}^6$ scales as the product of
weights of the six perfect matchings $t_1^4t_2^2$. This gives the CY condition $4b_1+2b_2 = 2$
and the volume function becomes
\beq
V \left ( b_1, b_2; \widetilde{SPP}_{\{1,-1,0\}} \right ) = \frac{1}{4b_1^2 b_2(2b_1+b_2)} .
\eeq
This gives a minimum at $b_1 = b_2 = 1/3$, reproducing the mesonic spectrum of $D_3$. Using this
result we can rewrite the two $U(1)$ fugacities as $t_1=t q, t_2 = t/q^2$, where $t$ is the
fugacity for the $R$ charge and $q$ is a fugacity for a global charge. The Hilbert series of
$D_3$ takes the form
\beq
g \left (t, q , x, y; \widetilde{SPP}_{\{1,-1,0\}} \right ) =\frac{1-t^6}{(1-t^2 q^2 x)(1-t^2 q^2 /x)(1-t^3 y)(1-t^3/y)(1-t^2/q^4)} .
\eeq
By symmetry the six normalized volumes are equal to $1/3$ and this gives R-charge $1/3$ for the
$B$ fields and R-charge $2/3$ for the $A$ and $C$ fields.

\subsubsection {(ii) $(k_1, k_2, k_3) = (-2,1,1)$} 
\begin{figure}[ht]
\begin{center}
  \includegraphics[totalheight=2.8cm]{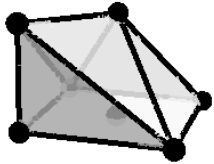}
  \caption{Toric diagram for $\widetilde{SPP}_{\{-2,1,1\}}$.}
  \label{spp211_toric}
\end{center}
\end{figure}

is interesting because it is the simplest case with irrational R-charges.
The two points $q_i$ split and we obtain a toric diagram with six external points as in
\fref{spp211_toric}.

The four-fold is obtained from the master space by modding out by $U(1)_2-U(1)_3$ acting on the
perfect matchings as $(1,-1,2,-2,0,0)$. By the redefinition $x_1\rightarrow x_1 w, x_2 = w^2$
this becomes equivalent to an integration over $w$.
\bea
&\qquad & \qquad\qquad\qquad\qquad\qquad\qquad\qquad g(t_1,t_2 ,,x_1,x_3; \widetilde{SPP}) = \\
&&\int  \frac{dw}{2\pi i w} \frac{1-t_1^4}{(1-t_1^2 x_1 x_3 w) (1-t_1^2 x_1 w/ x_3) (1-t_1^2
x_3/ x_1 w) (1-t_1^2/ x_1 x_3 w) (1-t_2 w^2)(1-t_2 / w^2)} \nonumber
\eea
This integral can be
easily done. We will not report the long resulting expression but we write the volume functional
to be minimized. The original top holomorphic form on $\mathbb{C}^6$ scales as the product of
weights of the six perfect matchings $t_1^4t_2^2$. In order to have a scaling of $2$ we write
$t_2=e^{-b \mu}$ and $t_1=e^{-(2-2 b)\mu/4}$. We can safely put $x_1=x_3=1$ by symmetry and we
have,
\beq
\lim_{\mu\rightarrow0} \mu^4 g \left (e^{-(1-b) \mu/2 }, e^{-\mu b},1,1 ;
\widetilde{SPP} \right ) =
\frac{16 -28 b +16 b^2 -3 b^3}{2 b (1-2 b +b^2)(16-32 b+24 b^2 -8 b^3 +b^4)} ,
\eeq
whose minimum is at
\beq
\label{rchargespp211}
b= \frac{1}{18}\left ( 19- \frac{37}{(431 -18\sqrt{417})^{1/3}} - (431-18\sqrt{417})^{1/3}\right ) \sim 0.319 .
\eeq
This is the R-charge of the fields called $B$. The R-charge for $A,C$  is then about $0.681$
corresponding to $(1-b)$.

Symmetry and the Hilbert series have determined uniquely the R-charges of the fields. There is a
baryonic symmetry that we can identify with $w$ but, by symmetry, it does not contribute to
R-charges.  It is nevertheless interesting to look at divisors and volumes. We have six external
points in the toric diagram, labeled by the six perfect matchings, corresponding to five-cycles
where we can wrap five-branes. Call $a_i$ the R-charge of a brane wrapped on the $i$-th cycle.
The six numbers $a_1,\ldots, a_6$ are in correspondence with $p_1,p_2,p_3, p_4,q_1,q_2$,
respectively. We can compute the value of $a_i$ by any of the methods in the appendix. Using the
Hilbert series method we have to recompute the Molien integral in (\ref{hsSPP}) with the
insertion of (the inverse of) the weight of the corresponding perfect matching. For example,
after all redefinitions, $p_1$ has weight $t_1 x_1 w z$ under the four toric symmetries, the
baryonic symmetry and the symplectic quotient charge $z$, so we compute
\bea
& \qquad &  \qquad \qquad \qquad \qquad \qquad \qquad  g(D_1; t_1  ,x_1,x_2,t;\widetilde{SPP}_{\{-2,1,1\}}) =\\
&&\int   \frac{dz}{2\pi i z} \frac{dw}{2\pi i w} \frac{(t_1 x_1 w z)^{-1}}{(1-t_1 x_1 w z)(1-t_1 z /x_1 w)(1-t_1 x_2 w^2/z)(1-t_1 w^2 /x_2 z)(1-t_3)(1-t_4)}
\nonumber\eea
and
\beq a_1 = \lim_{\mu\rightarrow0}\frac{1}{\mu}\left ( \frac{g(D_1; e^{-(1-b) \mu/2 }, e^{-\mu b} ,1,1;
\widetilde{SPP}) }{g( e^{-(1-b) \mu/2 }, e^{-\mu b },1,1;
\widetilde{SPP} )} - 1\right )  \sim 0.305 .
\eeq
Analogously we compute $a_2=a_1$, $a_3=a_4 \sim 0.319$ and $a_5=a_6 \sim 0.376$. The fields have
R-charges that follows from their expression in terms of perfect matchings given in Table
\ref{spptab},
\beq
A_1 \rightarrow a_1 +a_5\, , A_2 \rightarrow a_2 +a_6\, ,C_1 \rightarrow a_1 +a_6\, , C_2 \rightarrow a_2 +a_5\, , B_1 \rightarrow a_3\, , B_4 \rightarrow a_4 ,
\eeq
and we recover the previous result \eref{rchargespp211}. It is interesting that $a_{1,2}$ and
$a_{5,6}$ are different. We see that consistently with symmetry, the points are paired two by
two, but while $a_{3,4}$ are the R-charge values of the fields $B_{1,2}$, the R-charges
$a_{1,2,5,6}$ do not correspond to elementary fields. $A$ and $C$ can be obtained by wrapping
branes on pairs of cycles. It remains to understand whether there are consistent objects wrapped
on the cycles $1,2,5,6$ in the spectrum of the Chern-Simons theory.

\subsubsection {(iii) $(k_1, k_2, k_3) = (0,-1,1)$} was already worked out in detail in
\cite{Imamura:2008nn,Hanany:2008cd}. For completeness we briefly discuss this case as well. The
choice $(k_1, k_2, k_3) = (0,-1,1)$ gives the toric diagram of $\IC^2 / \IZ_2 \times  \IC^2$. It
is obtained by  dividing the master space by the gauge group without CS term, $U(1)_1$. As seen
from the previous assignment of charges, this is equivalent to integrate over $x_3$. The four
fugacities $t_1, t_2, x_1, x_2$ correspond to the four toric actions of the Calabi-Yau
four-fold. This computation is by now familiar and we obtain
\beq g(t_i,x_1;\widetilde{SPP}_{0,-1,1}) = \frac{1+t_1^4}{(1- t_1^4 x_1^2)(1-t_1^4/x_1^2)(1-t_2 x_2)(1-t_2 /
x_2)}.
\eeq
Thus, we find indeed the Hilbert series\footnote{As already discussed in
\cite{Imamura:2008nn,Hanany:2008cd}, the result can be easily   recovered by looking at
independent invariants under the $U(1)_1$  action: $M_{11} = A_1 A_2, M_{21} = C_1 A_1,
M_{12}=A_2 C_2, M_{22}=C_1 C_2, B_1,B_2$ which satisfy the  equations $M_{12} M_{21} = M_{11}
M_{22} = M_{11}^2$.} for $\IC^2 / \IZ_2 \times  \IC^2$.

By symmetry, we expect the same R-charge for the fields $A,B,C$ which should be $1/2$ to fit
with the superpotential. In fact, the minimization of the volume functional reproduces this
result. There is a baryonic charge but its value is by symmetry zero. The four external perfect
matchings $p_1,..,p_4$ in the 3d toric diagram  carry an R-charge $a_i$ that can be computed as
in the previous subsection, but with no surprises this time: $a_i=1/2$, consistent with our
previous discussion.




\subsection{The $\widetilde{\IC^3/\IZ_3}$ theories}

\begin{figure}[ht]
\begin{center}
  \hskip -6cm
  \includegraphics[totalheight=4.5cm]{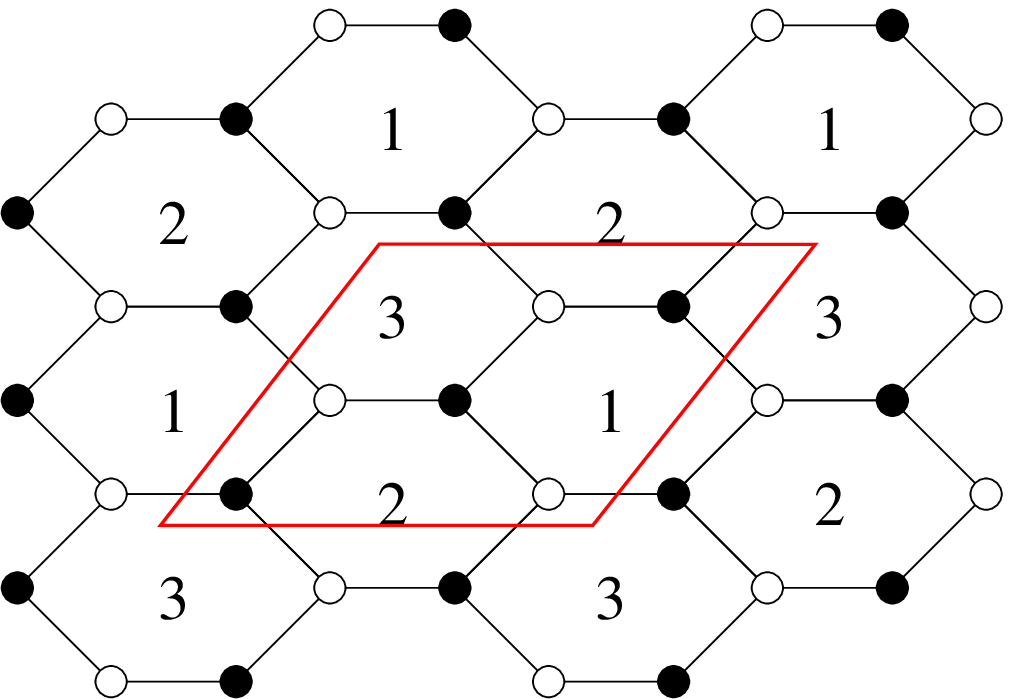}
  \vskip -4.0cm
  \hskip 8cm
  \includegraphics[totalheight=4.0cm]{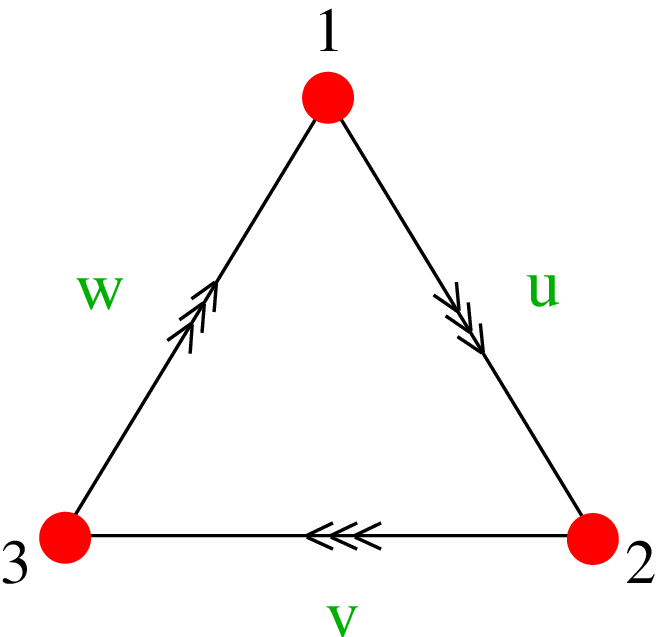}
  \vskip 0.5cm
  \caption{(i) Brane tiling for $\widetilde{\IC^3/\IZ_3}$ \ (ii) The corresponding quiver}
  \label{dp0}
\end{center}
\end{figure}

This set of models are given by the tiling and quiver of \fref{dp0}. There are three groups and
three sets of chiral fields $U_i,V_i,W_i, \, i=1,2,3$ transforming in the $(N,\bar N,0),
(0,N,\bar N)$ and $(\bar N,0, N)$ representation of the gauge group, respectively, and
interacting with the superpotential

\beq W= \epsilon_{ijk} U_i V_j W_k . \eeq

The levels of the groups are can be chosen to be $k_1, k_2$ and $-k_1-k_2$, respectively. These
models are special since they appear to be the simplest chiral (in the 3+1 dimensional sense)
models that exhibit a spectrum of non-trivial anomalous dimensions. The superpotential is cubic
and therefore allows for scaling dimensions which are far from the canonical scaling in 2+1
dimensions.

The Kasteleyn matrix is
\be
K = \left(
\begin{array}{ccc}
 z^{-{k_1}+{k_2}} & \ z^{-{k_1}-2 {k_2}} x & \ y^{-1} z^{2 {k_1}+{k_2}} \\
 z^{-{k_1}-2 {k_2}} & \ z^{2 {k_1}+{k_2}} & \ z^{-{k_1}+{k_2}} \\
 z^{2 {k_1}+{k_2}} y & \ z^{-{k_1}+{k_2}} & \ x^{-1} z^{-{k_1}-2 {k_2}}
\end{array}
\right) ,
\ee
and the permanent is
\be
  \textrm{perm} \, K = z^{-3 {k_1}-6 {k_2}}+z^{-3 {k_1}+3 {k_2}}+z^{6 {k_1}+3
{k_2}}+x^{-1}+y^{-1}+x y .
\ee
After an appropriate rescaling, this gives a toric diagram with the following points:

\vskip 0.2cm $(-1,0,0), \ (0,-1,0), \ (1,1,0), \ (0,0,-k_1-2k_2), \ (0,0,-k_1+k_2), \
(0,0,2k_1+k_2)$. \vskip 0.2cm

 \noindent These models are identified with the two parameter set
of theories\footnote{See \cite{Martelli:2008si} and the revised version of
\cite{Martelli:2008rt}.}, $Y^{p,k}(\IC P^2)$, with $p = k_1+k_2, k = 2 k_1 +  k_2$ for
$k_1,k_2\ge 0$. In the following, we will look at special cases.

\subsubsection{The cone over $M^{1,1,1}$}

The case of CS parameters  $k_1=k_2=1$ deserves special attention. The corresponding four-fold
is the cone over the coset manifold $M^{1,1,1}=SU(3)\times SU(2)\times U(1)/SU(2)\times U(1)
\times U(1)$ \cite{Castellani:1983mf} \footnote{This is called $M^{3,2}$ in
\cite{Martelli:2008rt}.} with global symmetry  $SU(3)\times SU(2)\times U(1)_R$. The coincidence
of this global symmetry with the gauge group of the standard model for particle interactions was
a reason for enhanced activity back in the 80's. An attempt to give a Yang-Mills theory dual can
be found in \cite{Fabbri:1999hw}. Here we focus on the Chern-Simons dual theory.

We can use the refined Kasteleyn matrix to learn about fields and perfect matchings
\beq
K = \left(
\begin{array}{ccc}
U_1 z^{-{k_1}+{k_2}} & V_2 \ z^{-{k_1}-2 {k_2}} x & W_3\ y^{-1} z^{2 {k_1}+{k_2}} \\
V_3 z^{-{k_1}-2 {k_2}} & W_1 \ z^{2 {k_1}+{k_2}} & U_2\ z^{-{k_1}+{k_2}} \\
W_2 z^{2 {k_1}+{k_2}} y & U_3\ z^{-{k_1}+{k_2}} & V_1\ x^{-1} z^{-{k_1}-2 {k_2}} ,
\end{array}
\right)
\eeq

\bea
  \textrm{perm} \, K = z^{-3 {k_1}-6 {k_2}} V_1 V_2 V_3+z^{-3 {k_1}+3 {k_2}} U_1 U_2 U_3+z^{6 {k_1}+3
{k_2}} W_1 W_2 W_3+x^{-1} U_1 V_1 W_1 \nonumber \\
+y^{-1} U_3 V_3 W_3+x y U_2 V_2 W_2 .
\eea
As we see, there are six perfect matchings $q_2,q_1,q_3,p_1,p_3,p_2$ corresponding with the
monomial in ${\rm perm} K$ with the given order. From the 3+1 dimensional perspective, the three
$p_i$ are external perfect matchings and the three $q_i$ are associated with the internal point
of multiplicity 3. From the 2+1 dimensional perspective, $q_2$ and $q_3$ become new external
points while $q_1$ remains an internal point. The five external perfect matchings $p_i,q_{2,3}$
can be set in correspondence with the five external points in the toric diagram and the
corresponding five divisors $D_i$. From the Kasteleyn matrix we see that $U_1$ belongs to the
perfect matchings $p_1$ and $q_1$ so it can be parametrized as the product $p_1q_1$. in a
similar fashion we read the parameterization of the other fields in terms of perfect matchings,
\beq
U_i = p_i q_1 ,\qquad V_i = p_i q_2 , \qquad W_i = p_i q_3 \label{pm}
\eeq

The moduli space is given by modding out the five dimensional master space by the $U(1)$ gauge
symmetry prescribed by the CS terms. Let us describe both the master space and the resulting
four-dimensional Calabi-Yau.

The master space is given by the perfect matchings modulo relations. There is one  relation
among perfect matching $p_1+p_2+p_3 =q_1+q_2+q_3$ and this gives the, by now usual
\cite{Forcella:2008bb}, description of the master space as $\mathbb{C}^6/\{-1,-1,-1,1,1,1\}$,
where we order the perfect matchings as $p_1,p_2,p_3,q_1,q_2,q_3$. The master space is a five
dimensional toric variety with $SU(3)\times SU(3)\times U(1)$ symmetry, where the second $SU(3)$
is hidden from a 3+1 dimensional perspective. We can introduce weights for the action of the
global symmetry on perfect matchings as follows:
\beq p_1,p_2,p_3,q_1,q_2,q_3 \,\, \rightarrow\,\, t y, t x,  t/xy, x_1,  x_2, 1/x_2 x_1\eeq
where $t$ is the $U(1)$ charge, $x,y$ are weights for the first  $SU(3)$ and $x_1,x_2$ are weights for the second  $SU(3)$. As seen from  formula (\ref{pm}) $x_1$ and $x_2$ correspond to the two independent charges under the gauge group. The Hilbert series for the master space is
\beq
\int \frac{dz}{2\pi i z} \frac{1}{(1-t y/z)(1-t x/z)(1-t x y/z)(1-x_1 z)(1-z x_2)(1-z/ x_2 x_1)}
\eeq

The Calabi Yau four-fold is obtained now by modding by the gauge group $U(1)_1-U(1)_2$ which, as
seen from equation (\ref{pm}) corresponds to  the action $\{0,0,0,2,-1,-1\}$ on perfect
matchings and it breaks the global symmetry to $SU(3)\times SU(2)\times U(1)$. Note that part of
the hidden symmetry now becomes a symmetry of the mesonic moduli space in the 2+1 dimensional
theory. By redefining $x_1=w^2, x_2=\tilde x/w$, this just corresponds to integrating over $w$.
We can use $t,x,y,\tilde x$ to parametrize the four toric symmetries of the Calabi-Yau. $\tilde
x$ is now interpreted as an $SU(2)$ weight. The Hilbert series for the mesonic moduli space
depends on $t,x,y,\tilde x$ and is given by
\bea
g(t,x,y,\tilde x; \widetilde {\mathbb{C}^3/Z_3)}& = & \int \frac{dz}{2\pi i z}\frac{dw} {2\pi i } \frac{1}{((1-t y/z)(1-t x/z)(1-t x y/z)(1-w^2 z)(1-z \tilde x/w)(1-z/ w \tilde x))} \nonumber\\
&& = \sum_{k=0}^\infty [3k,0; 2k] t^{3k}
\eea
where $[n,m;s]$ denotes irreps of $SU(3)\times SU(2)$. From the last expression we recognize
indeed the KK spectrum of M theory compactified on  $M^{1,1,1}$ \cite{Fabbri:1999hw}.

We can extract the volume formula from the Hilbert series expression and minimize it. However we
know the result without need of computation. The $SU(3)\times SU(2)$ symmetry immediately gives
$x=y=\tilde x=1$. Moreover the Calabi-Yau  top form scale as $\prod p_i q_i\sim t^3$ and
therefore $t$ corresponds to a dimension $2/3$. The see that the KK spectrum consists of
multiplets of dimensions $2k, k=0, 1, 2, \ldots$ as indeed known from supergravity
\cite{Fabbri:1999mk,Fabbri:1999hw}.

The full moduli space, including baryonic operators, has an extra charge with fugacity $w$ as
above. The mesonic operators are independent of $w$ but the R-charges of the elementary fields
depend on $w$. Since $SU(2)$ exchanges $V,W$, they have same R-charge but this can be different
from the R-charge of $U$. We can use the external perfect matchings to parametrize the five
global charges. Introducing R-charges $a_1,a_2,a_3,a_4,a_5$ associated with
$p_1,p_2,p_3,q_2,q_3$, the fields have R-charge:
\beq
U_i \rightarrow a_i, \qquad V_i \rightarrow a_i+a_4 ,\qquad W_i \rightarrow a_i+a_5 \qquad i=1,2,3
\eeq
It is tempting to compute the numbers $a_i$ with the same rule as in 3+1 dimensions: $a_i$ is
the R-charge corresponding to a brane wrapped on the five-cycle corresponding to the $i$-th
external point of the toric diagram and it is computed as a normalized volume.

The volumes  were computed in \cite{Fabbri:1999hw}  and reobtained in \cite{Martelli:2008si}.
Divisors $D_{1,2,3}$ correspond to a normalized volume $4/9$ and divisors $D_{4,5}$ to $1/3$. We
can recheck it with the method discussed in the Appendix which indeed gives, for example,
\beq
g(D_1,; t,x,y,\tilde x ; \widetilde{\mathbb{C}^3/Z_3})=\int \frac{dz}{2\pi i  z }\frac{dw}{2\pi i w} \frac{(t y/z)^{-1}}{((1-t y/z)(1-t x/z)(1-t x y)(1-w^2
z)(1-z \tilde x/w)(1-z/ w \tilde x))}
\eeq
and
\beq
a_1 = \lim_{\mu\rightarrow0}\frac{1}{\mu}\left ( \frac{g(D_1; e^{-2 \mu/3 },1,1, 1;
\widetilde{\mathbb{C}^3/Z_3}) }{g(e^{-2\mu /3},1,1,1 ;\widetilde{\mathbb{C}^3/Z_3})} - 1\right )  = \frac{4}{9} \eeq
and similarly for the other divisors.

We get $a_1=a_2=a_3=4/9$ and $a_4=a_5=1/3$. We see from (\ref{pm}) that we expect R-charge $1/3$
for $U$ and $7/9$ for $V,W$. This is consistent with the CS theory lagrangian. As in the other
examples in this paper we see that no elementary field is associated with $D_{4,5}$ and the
value $4/9$. It remains an open question to understand whether there are really states in the
spectrum obtained from M5 branes wrapped on the base of $D_4$ and $D_5$ and what is their
description in the CS theory.

\begin{figure}[ht]
\begin{center}
  \includegraphics[totalheight=3.0cm]{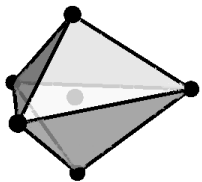}
  \caption{ Toric diagram for the cone over $M^{1,1,1}$.}
  \label{m32_3d}
\end{center}
\end{figure}

\subsubsection{The case of $\widetilde {\mathbb{C}^3/\IZ_3}_{\{1,0,-1\}}$}
This case corresponds to the manifold $Y^{12}(\mathbb{C}P_2)$ \cite{Martelli:2008rt}. All the
points $q_i$ now become external points. There are altogether five external points because $q_1$
and $q_2$ coincide.

The Calabi Yau four-fold is now obtained by modding out the master space by the gauge group
$U(1)_2$ with action $\{0,0,0,-1,1,0\}$ on the perfect matchings. This breaks the global
symmetry to $SU(3)\times U(1)^2$. After the redefinition $x_1=1/w,x_2=w/q$  it just corresponds
to integrate over $w$. We can still use $t, x, y, q$ to parametrize the four toric symmetries of
the Calabi-Yau. The Hilbert series for the mesonic moduli space depends on $t, x, y, q$ and is
given by
\beq
g(t, x, y, q; \widetilde {\mathbb{C}^3/Z_3}_{\{1,0,-1\})}=  \int \frac{dz}{2\pi i z}\frac{dw} {2\pi i w} \frac{1}{((1-t y/z)(1-t x/z)(1-t x y/z)(1-  z/w)(1-z w / q)(1-z q))}
\eeq
The explicit expression is too long and we report it only in the case $x = y = q = 1$,
\beq
g(t,x=1,y=1,q=1; \widetilde {\mathbb{C}^3/Z_3}_{\{1,0,-1\}})  = \frac{1+2 t+6 t^2+2 t^3 +t^4}{(1-t)(1-t^2)^3} .
\eeq

We can extract the volume formula from the Hilbert series expression and minimize it. From the
scaling of the top holomorphic form as $t^3$ we still have that the charge for $t$ is $2/3$. The
$SU(3)$ symmetry immediately gives  $x=y=1$ and we obtain the volume function,
\beq
\lim_{\mu\rightarrow0} \mu^4 g \left (e^{-2 \mu/3 },1,1, e^{-\mu b_{q}};
\widetilde{\IC^3/\IZ_3}_{\{1,-1,0\}} \right ) = \frac{243 (16+12 b_{q} + 9 b_{q}^2)}{(8+6 b_{q} - 9 b_{q}^2)^3}
\eeq
whose minimization gives
\beq b_{q}= \frac{1}{6} \left ((181+24 \sqrt{78})^{1/3}-\frac{23}{ (181+24 \sqrt{78})^{1/3})} -3\right ) \sim 0.197 \eeq

The full moduli space, including baryonic operators, has an extra charge with fugacity $q$ as
above. We use again the external perfect matchings to parametrize the five global charges. This
time we introduce $a_1,a_2,a_3,a_4/2,a_5$ associated with $p_1,p_2,p_3,q_1=q_2,q_3$, so that the
fields have R-charge:
\beq U_i \rightarrow a_i +a_4/2,  \qquad V_i \rightarrow a_i+a_4/2 ,
\qquad       W_i \rightarrow a_i+a_5 \qquad i=1,2,3
\eeq
The values of $a_i$ can be computed as in the previous examples, obtaining rational expressions
in $b_{q}$ too long to be reported here. The only subtlety, compared with previous cases, is
that we have an external point with multiplicity two: the corresponding R-charge has to be
divided among the corresponding fields \footnote{A similar ambiguity appears in the computation
of $a_4$ using the Hilbert series for the line bundle $D_4$: we need to use the perfect matching
$q_1+q_2$.}.  The numerical value is:
\beq
a_1=a_2=a_3=0.451\, ,\qquad a_4=0.235\, \qquad a_5 =0.414
\eeq
The result seems to agree with the numerical computation in \cite{Martelli:2008rt} using the explicit metric.

We see from (\ref{pm}) that we expect R-charge $0.569$ for $U,V$ and $0.865$ for $W$. This is
consistent with the CS theory lagrangian. As in the other examples we see that no elementary
field is associated with $D_{4,5}$ and the values $a_4,a_5$.
\subsection{The $\tilde{\IF}_0$ theories}

\begin{figure}[ht]
\begin{center}
  \includegraphics[totalheight=4.0cm]{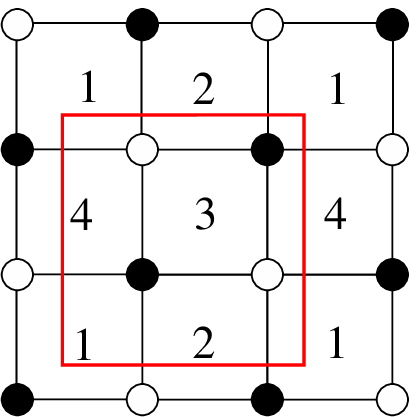}
  \caption{Brane tiling for $\IF_0^I$.}
  \label{f0_2}
\end{center}
\end{figure}

\begin{figure}[ht]
\begin{center}
  \includegraphics[totalheight=4.5cm]{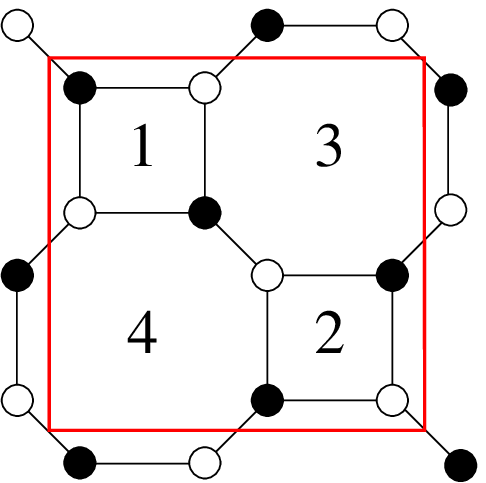}
  \caption{Brane tiling for $\IF_0^{II}$.
  }
  \label{f0}
\end{center}
\end{figure}

$\IF_0$ has two toric phases (in 3+1 dimensions) \cite{Feng:2000mi}. Their tilings are depicted
in \fref{f0_2}, where we denote the model as $\IF_0^I$, and in \fref{f0} where the model is
denoted by $\IF_0^{II}$. (This order of notation is chosen based on the complexity of the
models. For instance, the number of fields is 8 in $\IF_0^{I}$ and 12 in $\IF_0^{II}$.) A choice
of 2+1 dimensional CS levels does not in general commute with 3+1 dimensional toric duality.
Therefore, we need to treat each model separately: they give rise to their own set of 2+1
dimensional field theories. It is essential to investigate the relation between toric duality
and the choice of CS levels but this will not be done in the present paper (see however some
speculations about a possible dual ABJM theory in Appendix \ref{app2}). We proceed as in the
previous examples.

The levels of the groups are chosen to be $k_1, k_2, k_3$ and $k_4=-k_1-k_2-k_3$, respectively.
We write the Kasteleyn matrix,

\be
K = \left(
\begin{array}{cc}
 z^{{k_3}-{k_2}}+z^{{k_1}-{k_4}} x & \ z^{-{k_3}+{k_4}}+\frac{z^{-{k_1}+{k_2}}}{y} \\
 z^{-{k_3}+{k_4}}+z^{-{k_1}+{k_2}} y & \ z^{{k_3}-{k_2}}+\frac{z^{{k_1}-{k_4}}}{x}
\end{array}
\right) ,
\ee

and find that after substituting $k_4=-k_1-k_2-k_3$, the permanent is
\bea
\textrm{perm} \, K  &=&  z^{-2 {k_1}-2 {k_2}-4 {k_3}}+z^{-2 {k_2}+2 {k_3}}+z^{-2 {k_1}+2
{k_2}}+z^{4 {k_1}+2 {k_2}+2 {k_3}} \\
 & & +\frac{z^{2 {k_1}+2 {k_3}}}{x}+z^{2 {k_1}+2 {k_3}} x+\frac{z^{-2 {k_1}-2 {k_3}}}{y}+z^{-2 {k_1}-2 {k_3}} y .
\eea

To disentangle this expression let us define linear combinations of levels, $a=k_1+k_3$,
$b=k_2+k_3$ and $c=k_1+k_2$, and write
\be
\textrm{perm} \, K  = \left(x+\frac{1}{x}\right) z^a + \left(y+\frac{1}{y}\right) z^{-a} +
z^{-a-b}+z^{-a+b} + z^{a-c} + z^{a+c} .
\ee
This gives a three-parameter set of toric moduli spaces.

By setting $a=1$ and $b=c=0$ (that is $k_1=-k_2=k_3$), we get $(\IC^2 / \IZ_2)^2$ as the moduli
space for the theory $\tilde{\IF}^{I}_{0 \{1,-1,1,-1\}}$. A similar analysis to the previous
examples can be done for scaling exponents and Hilbert series but we will skip this.


\begin{figure}[ht]
\begin{center}
  \includegraphics[totalheight=2.5cm]{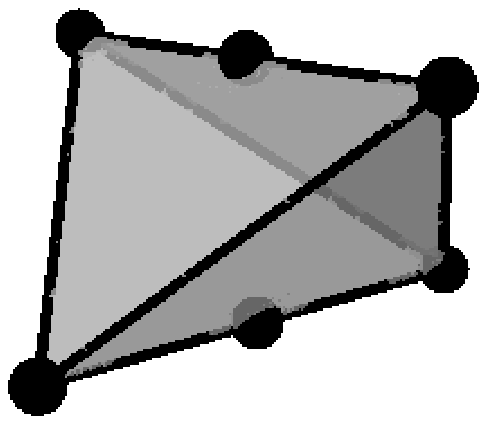}
  \caption{ 3d toric diagram for $(\IC^2 / \IZ_2)^2$.}
  \label{c2z2_2_3d}
\end{center}
\end{figure}

\subsubsection{The $\IF_0^{II}$ tiling and its family of theories}

For model $\IF_0^{II}$ we will be very brief. Set the levels of the groups as $k_1, k_2, k_3$
and $k_4=-k_1-k_2-k_3$, respectively.

The resulting Kasteleyn matrix and permanent are

\be K =\left(
\begin{array}{cccc}
 z^{-{k_1}+{k_4}} & z^{{k_1}-{k_3}} & 0 & \frac{z^{{k_3}-{k_4}}}{x y} \\
 z^{{k_1}-{k_3}} & z^{-{k_1}+{k_4}} & z^{{k_3}-{k_4}} & 0 \\
 0 & z^{{k_3}-{k_4}} x & z^{{k_2}-{k_3}} & z^{-{k_2}+{k_4}} \\
 z^{{k_3}-{k_4}} y & 0 & z^{-{k_2}+{k_4}} & z^{{k_2}-{k_3}}
\end{array}
\right)
\ee


\bea
t^{4 {k_3}} \cdot \textrm{perm} \, K  = t^{-4 {k_1}}+t^{-4 {k_2}}+t^{-6
({k_1}+{k_2})}+t^{2 ({k_1}+{k_2})}+t^{4 ({k_1}+{k_2}+3 {k_3})} \\
+t^{2 ({k_2} +3 {k_3})} \left(\frac{1}{x}+x\right)+t^{2 ({k_1}+3 {k_3})}
\left(\frac{1}{y}+y\right)
\eea


In the case of $k_1 = k_2$, where the nodes that are external points in the 2d toric diagram are
still coplanar, the toric diagram obtained from the above permanent seems to match a subset of
diagrams of \cite{Martelli:2008si} for $Y^{p,k}(\IC P^1 \times \IC P^1)$.


\section{Conclusions}
\label{section_conclusions}

In this paper we study in detail the properties of 2+1 dimensional Chern-Simons theories with an
abelian moduli space that is a Calabi-Yau four-fold following the construction in
\cite{Hanany:2008cd}. In particular, we describe an extension of the ``fast forward algorithm''
of \cite{Franco:2005rj} to efficiently determine the toric data of the Calabi-Yau starting from
the tiling.

Such theories stand naturally as candidates for the world-volume theory of membranes probing the
Sasaki-Einstein base of the Calabi-Yau and they may become an important ingredient in our
understanding of the AdS$_4$/CFT$_3$ correspondence. In order to check the duality between the
Chern-Simons theory and the $AdS_4\times H$ background, we study the spectrum of mesonic and
baryonic excitations, as predicted by the supergravity dual. Unfortunately, it is still
difficult to have exact quantum field theory results in 2+1 dimensions, so we can only perform
some consistency checks on our theories. As expected, the mesonic spectrum of a Chern-Simons
theory agrees with the KK spectrum of the compactification on $H$. This is not a surprise: as is
well known from similar analysis in 3+1 dimensions, this is a consequence of the relation
between holomorphic functions on the Calabi-Yau and eigenvectors of the Laplacian on $H$. The
R-charges that we obtain by minimization are consistent in all our examples with the
superpotential structure of the Chern-Simons Lagrangian and satisfy all the relevant unitary
bounds. We point out a puzzle regarding the baryonic spectrum, where it seems that five-branes
wrapped on certain cycles do not correspond to baryons made with elementary fields. This remains
a problem to be solved for a proper understanding of the proposed duality.

The class of Chern-Simons theories obtained by tilings probably does not exhaust all the
theories dual to Calabi-Yau four-folds, not even the toric ones. The analysis of the moduli
space is not limited to theories coming from tilings and could be applied to more general
theories, obtained for example from crystals \cite{Lee:2006hw,Lee:2007kv}. This is left for
future work.

\vskip 0.5cm

{\bf Acknowledgements}

A.~H.~ would like to thank the ULB and VUB physics departments of Brussels for kind hospitality
during the completion of this project. D.~V.~ thanks John McGreevy and Alessandro Tomasiello for
discussions. This work is supported in part by funds provided by the U.S. Department of Energy
(D.O.E.) under cooperative research agreement DE-FG0205ER41360. A.~Z.~ is supported in part by
INFN  and by the European Community's Human Potential Program MRTN-CT-2004-005104.

\appendix
\section{Appendix: Prescriptions for computing the Hilbert series}
\label{prehS}

Consider the description of the Calabi-Yau four-fold $X$ as a symplectic quotient on the space
of perfect matchings. We usually want to write refined Hilbert series depending on a set of
global charges characterizing the four-fold. In particular we always introduce at least four
weights corresponding to the four toric $U(1)$ actions on $X$.

The perfect matchings  are indicated as $p_\alpha$ where the index $\alpha$ runs from one to the
number $c$ of integers points in the 2d toric diagram, including multiplicities. The number $c$
can be large and depends in a non trivial way on the form of the tiling. The perfect matchings
are subsets of the collections of edges (elementary fields) and we can form formal linear
combinations with integer coefficients $\sum n_\alpha p_\alpha,\, n_\alpha \in \mathbb{Z}$.
Please notice that we will use an additive notation for perfect matchings, when regarded as
collections of fields. The $p_\alpha$, as collections of edges in the tiling,  will in general
satisfy some linear relations $\sum_\alpha Q_\alpha p_\alpha = 0$ that translate into vectors of
charges for a symplectic quotient. The coherent component of the master space, which is the
biggest irreducible component of the F-term solutions, has dimension $g+2$ and it is then
obtained by modding the space of perfect matchings $\mathbb{C}^c$ by the charge vectors
corresponding to the $c-g-2$  linear relations satisfied by the perfect matchings. Another way
to say this is that the solution of the F-terms can be written as
\begin{equation}
X_i = \prod_{\alpha = 1}^c p_\alpha^{P_{i \alpha}} \ ,
\label{pmpar2}
\end{equation}
where the element of the matrix $P_{i \alpha}$ is $1$ if the field $X_i$ belongs to the perfect
matching $p_\alpha$ and $0$ if it does not. The redundancy in the parametrization (\ref{pmpar2})
is given by the kernel of the matrix $P$ and one can prove that the vectors of charges $Q$  span
precisely the kernel of $P$. All the global and gauge symmetries of the CS theory act on the
solutions of F-terms and can be lifted to an action on $p_\alpha$. We can then give weights
under global and gauge actions to the $p_\alpha$.

We deal in the text with many different symplectic quotients of $\mathbb{C}^c$:

\begin{itemize}
\item By modding by the $\mathbb{C}^*$ action corresponding to linear relations
among perfect matchings, we obtain the master space.
\item By adding the $g-2$ gauge charges we obtain the CY four-fold $X$. 
\item By adding the last  gauge charge that is reduced to discrete symmetry by the presence of the CS terms we obtain the CY three-fold associated with the tiling.
\end{itemize}

The Hilbert series for all these cases can be computed with the integral Molien formula which
schematically reads
\beq g(t_i; {X})=\int \prod_{i=1}^{G} \frac{dz_i}{2\pi i z_i} \frac{1}{\prod_{\alpha=1}^{c} (1 - t_\alpha Z_\alpha )} \eeq
where $G$ is the total number of $\mathbb{C}^*$ actions we are modding out. In this formula the
dummy variable $t_\alpha$ correspond to the perfect matching $p_\alpha$ and it is the weight of
the perfect matching itself under the global symmetry of the theory. $Z_\alpha$ denotes the
monomial weight of the $\alpha$-th perfect matching $p_\alpha$ in terms of the $z_\alpha$. For
example, if we want the Hilbert series of the Calabi-Yau four-fold $X$, $t_\alpha$ will denote
the weight under the four global charges of the toric action on $X$ and $Z_\alpha$ is the weight
under the a total of $c-4$ charges, divided into $c-g-2$ charges coming from the linear
relations among perfect matchings and the $g-2$ D-term  charges. Notice that we can assign $c$
different weights $t_\alpha$ to the perfect matchings, but, due to the $c-4$ integrations, only
four of them will be independent. Similar arguments apply to the computation of the Hilbert
series for the master space or for the Calabi-Yau three-fold associated with the tiling.

Sometimes the description in terms of master space, although conceptually crystal clear and
directly connected to the field theory, can be cumbersome because of the many contour
integrations to be performed, especially when $c$ is large. We can then resort to different
types of descriptions. Any toric variety can be written, for example, as a symplectic quotient
in $\mathbb{C}^d$ where $d$ is the number of external points in the toric diagram \cite{fulton}.
The information about the 3d toric diagram comes to us from the magnetic Kasteleyn matrix and
allows for computation using a Molien integral on $d$ variables and $d-4$ contour integrals.
This description is useful when $d \ll c$.  Finally, in some lucky cases, we will be able to
write our variety, or the master space, as a set of algebraic equations which defines a complete
intersection. The computation of the Hilbert series is then straightforward due to this
property.

To study volumes, we can take two different approaches.  Following  \cite{Butti:2006au}, we are
led to investigate divisors on the CY and their associated line bundles. Recall that there is a
correspondence between external points of the toric diagram and divisors $D_a$. We call external
perfect matchings those corresponding to external points in the toric diagram. The Hilbert
series for the line bundle associated with $D_a$ is obtained as follows. Given a divisor $D_a$
corresponding to an external perfect matching $p_a$, we can conjecture the following modified
Molien formula for the Hilbert series of holomorphic sections of $D_a$
\beq
g(D_a; {X}) = \int \prod_{i=1}^{G} \frac{dz_i}{2\pi i z_i} \frac{(t_a Z_a)^{-1} }{\prod_{\alpha=1}^{c} (1 - t_\alpha Z_\alpha)}
\label{divisors}
\eeq
The rationale for this formula comes from the fact that every line bundle on the CY four-fold
should come from the ambient space $\mathbb{C}^c$. A line bundle on  $\mathbb{C}^c$ is
necessarily free, but we can obtain line bundle on the CY by letting the charges $z_i$ to act on
the fiber.

As discussed in section \ref{hilbert} we can extract the volume of the base of $X$ from the
leading pole of the Hilbert series for $X$. We can analogously extract the volumes of the base
of $D_a$ from the Hilbert series for holomorphic sections of the line bundle
\cite{Butti:2006au}. All these results will depend on a Reeb vector $b=(b_1,b_2,b_3,b_4)$,
specifying a linear combination of the four toric actions, restricted by the only condition that
it gives charges $2$ to the holomorphic top form. The actual value of $b$ is obtained by
minimizing the volume of the base of $X$. Knowing $b$ at the minimum we can compute volumes of
the seven manifold $X$ and its five-cycles.

Since we are interested in the R-charge of a brane wrapped on the base of the divisor, we need
to normalize the volumes according to formula (\ref{Rc}). The following nice formula works both
in 3+1 dimensions and in 2+1 dimensions,
\beq
\frac{g(D_i, e^{-\mu b_i}; {X})}{g(e^{-\mu b_i}; {X})}\sim 1 + \Delta (b_i) \mu + \ldots
\label{Rcharges}
\eeq
Here $b$ is the Reeb vector whose value is obtained by minimizing the coefficient of the leading
pole of the Hilbert series $g(t_i; {X})$ for $\mu\rightarrow 0$ as discussed in the main text.
We point out that the previous formulae stand as conjectures tested in many interesting cases
against known results. In particular they should be used with care in cases where the Calabi-Yau
has singularities or multiplicities on the external points.

By re-elaborating the results in \cite{Martelli:2005tp} we can also write an explicit formula in
terms of the toric data. Given an external point $v_a$ of the 2d toric diagram, with associated
divisors $D_a$, consider the clockwise ordered sequence of vectors $w_k, k=1,\ldots , n_a$ in
the toric diagram that are adjacent to $v_a$. Define
\beq
F_a = \sum_{k=2}^{n_{a}-1}  \frac{
(v_i,w_{k-1},w_k,w_{k_1})(v_i,w_k,w_1,w_{n_a})}{(v_i,b,w_k,w_{k+1})
(v_i,b,w_{k-1},w_k)(v_i,b,w_1,w_{n_a})}
\label{MSvol}
\eeq
where $(V_1,V_2,V_3,V_4)$ denotes the determinant of four vectors $V_{1,2,3,4}$. The expression
for the R-charge of a five-brane wrapped on the base of $D_a$  is then
\beq R_a= \frac{2 F_a}{\sum_a F_a} \eeq
These formulae depend on the Reeb vector $b$, whose value can be found by minimizing the volume
functional which in the new setting is given by  $\sum_a F_a$. $b$ is always constrained by the
request that the holomorphic top form scales appropriately; in the approach with toric data and
in in all our examples where the CY condition is enforced by taking vectors $v_a$ with fourth
coordinates equal to $1$, this means $b_4=4$. Due to different notations, the vector $b$
entering in these equations is in general related to the vector $b$ entering the Molien formula
by a change of basis.

In this approach it is easy to see the important fact that the normalized volumes always add up
to $2$.

We now give an explicit example of the use of the previous formula. Many other examples are
scattered along the paper.

\subsection{The cone over  $Q^{1,1,1}$}

We use an  example where all about Hilbert series and volumes is under control, the cone over
the manifold $Q^{1,1,1}= SU(2)^3/U(1)\times U(1)$ with symmetry $SU(2)^3\times U(1)_R$ . It is
not clear at the moment how to construct a Chern-Simons theory with non-abelian moduli space the
symmetric product of copies of $C(Q^{1,1,1})$ \footnote{For Yang-Mills approaches see
\cite{Fabbri:1999hw,Oh:1998qi}.}. Nevertheless, it makes perfect sense to analyze the Hilbert
series and the volume of divisors as a check of our general formulas. Most of the results could
be predicted by using symmetries only, but the purpose of the following discussion is to explain
how to use the Hilbert series and the toric data to determine the Reeb vector and volumes.

The nice thing is that the cone over $Q^{1,1,1}$ has a symplectic quotient description:
$C(Q^{1,1,1})$ is $\mathbb{C}^6$ modded by two $\mathbb{C}^*$ with charge vectors
$\{1,1,-1,-1,0,0\},\{0,0,1,1,-1,-1\}$. The toric diagram has indeed six external points
satisfying two linear relations given by the previous vectors.
\begin{figure}[ht]
\begin{center}
  \includegraphics[totalheight=3.0cm]{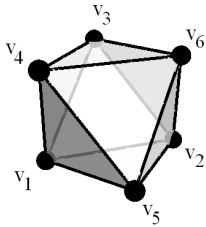}
  \caption{ Toric diagram for $Q^{1,1,1}$.}
  \label{q111td}
\end{center}
\end{figure}

Introduce weights $t_1=tx_1,t_2=t/x_1,t_3=t x_2,t_4=t/x_2, t_5=t x_3,t_6=t/ x_3$ for the 6 external points, where $x_1, x_2, x_3$ are weights for $SU(2)^3$ and $t$ is a weight for $U(1)_R$. The Hilbert series is
\bea
\nonumber
g(t_i; C(Q^{1,1,1}))&=&\int \frac{dz_1}{2\pi i z_1} \frac{dz_2}{2\pi i z_2}\frac{1}{(1-t_1 z_1)(1-t_2 z_1)(1-t_3 z_2 / z_1)(1-t_4 z_2 / z_1)(1-t_5/ z_2)(1-t_6/ z_2)} \\ &=& \sum_{n=0}^\infty [n;n;n] t^{3 n} ,
\eea
where $[n;n;n]$ denotes the representation of dimension $(n+1)^3$ of $SU(2)^3$. We recognize the
familiar KK spectrum of chiral multiplets in the compactification on $Q^{1,1,1}$, which indeed
transform in the $[n;n;n]$ representation \cite{Fabbri:1999mk}.

{As an aside we note that explicit expressions for the characters of $SU(2)$ are simple and take the form,
\beq
[n_1; n_2; n_3] = \prod_{i=1}^3 \frac{x_i^{n_i+1}-x_i^{-n_i-1}}{x_i-x_i^{-1}} .
\eeq}
Using this expression the Hilbert series takes an alternative palindromic form
\beq
g \left (t, x_1, x_2, x_3; C \left (Q^{1,1,1} \right ) \right ) = \left ( 1+t^{18} - \left ( [2; 0; 0] + [0; 2; 0] + [0; 0; 2] \right ) \left ( t^6 + t^{12} \right ) + 2 [1; 1; 1] t^9 \right ) PE \left [ \left [ 1; 1; 1\right ] t^3 \right ] ,
\eeq
where this expression indicates that the generators of $C \left (Q^{1,1,1} \right )$ transform
in $[1; 1; 1]$ of $SU(2)^3$.

The volume function is obtained as the leading pole of the Hilbert series for $t_i\rightarrow
1$. We define $t=e^{- \mu b}, x_1=e^{- \mu b_1 }, x_2=e^{- \mu b_2}, x_3=e^{- \mu b_3}$ and take
the $\mu\rightarrow 0$ limit. The four parameters $b,b_1,b_2,b_3$ are not independent. They are
restricted by the request that the holomorphic top form scales with charge $2$. Since
holomorphic top form transform as $t_1\cdots t_6 = t^6$ we learn that $b=1/3$. The volume
functional is then
\beq
\lim_{\mu\rightarrow0} \mu^4 g \left (e^{-\mu/3},e^{- \mu b_1 },e^{- \mu b_2}, e^{-\mu b_3};
C(Q^{1,1,1}) \right )
\eeq
The result is a rational function of $b_1, b_2, b_3$ that should be minimized. Minimization of a
function of three variables can be a non trivial task. In the case of $C(Q^{1,1,1})$ we do not
really need to minimize because we know by symmetry that the result will be $b_1=b_2=b_3=0$. We
leave to the skeptic reader the evaluation and minimization of the previous quantity.

For the same reason, all divisors have the same volume. We sketch the computation of the
normalized volume with two different methods. With the first method, we compute the Hilbert
series that counts holomorphic sections of a line bundle. By symmetry we can choose any divisor.
Let us choose $D_1$ associated with the external point with weight $t_1$. We need to compute the
Molien integral with the insertion of the (inverse of) the weight of the external point under
all charges. The point $1$ has weight $t_1=t x_1$ and charge $z_1$ under the two symplectic
actions. We thus have, according to (\ref{divisors}),
\beq
g(D_1; t, x_i; C(Q^{1,1,1}))=\oint \frac{dz_1}{2\pi i z_1}  \frac{dz_2}{2\pi i z_2}\frac{(t_1 z_1)^{-1}}{(1-t_1 z_1)(1-t_2 z_1)(1-t_3 z_2 / z_1)(1-t_4 z_2 / z_1)(1-t_5 / z_2)(1-t_6 / z_2)}
\eeq
Notice that the insertion is just the first monomial appearing in the denominator.
Finally, using formula (\ref{Rcharges}) for the normalized volume we find
\beq \lim_{\mu\rightarrow0}\frac{1}{\mu}\left ( \frac{g(D_1; e^{- \mu/3 },1,1, 1;
C(Q^{1,1,1})) }{   g(D_1; e^{- \mu/3 },1,1, 1; C(Q^{1,1,1} ))} - 1\right ) = \frac{1}{3}\eeq
which is indeed the right result for volumes in $Q^{1,1,1}$ \cite{Fabbri:1999hw}. Obviously, due
to the high symmetry of the problem the outcome is just the scaling dimension of the only
non-trivial variable, $t$. We see in the main text many non-trivial applications of this method.

The second method uses the toric data and formula (\ref{MSvol}). The toric vectors for $C(Q^{1,1,1})$ are
\beq
v_1=(1,0,0,1)\,\,\, v_2=(0,1,0,1)\,\,\, v_3=(0,0,1,1)\,\,\, v_4=(1,0,1,1)\,\,\, v_5=(1,1,0,1)\,\,\, v_6=(0,1,1,1)
\eeq
We need to compute a volume, as a function of $b$, for each external point $v_a$. These volumes
are proportional to the quantities $F_a$ given in \eref{MSvol}. Each $F_a$ is a sum over
contributions coming from the points connected to $v_a$ by an edge. We need to specify an order
to use during the computation: the points are ordered clockwise as seen from the point $v_a$.
For example, we see from Figure \ref{q111td} that the point $1$ is connected to the points
$5,2,3,4$ with this particular order. We then apply formula (\ref{MSvol}) with $v_1=(1,0,0,1)$
and the string of points $w=\{v_5,v_2,v_3,v_4\}$ obtaining
\beq F_1=\frac{b_2+b_3}{b_2 b_3(b_1-b_4)(b_1+b_2+b_3-b_4)}
\eeq
and similar expressions for the other $F_a$. The volume functional is
$Z=F_1+F_2+F_3+F_4+F_5+F_6$ which should be minimized with the constraint that $b_4=4$. In this
example, by symmetry of the toric diagram, we can predict that $b_1=b_2=b_3\equiv B$ at the
minimum and we obtain the functional
\beq Z(B)= \frac{24}{B(B-4)(32-36 B+9 B^2)}\eeq
whose minimum is at $B=2$. We then have
\beq \frac{ 2 F_a}{\sum_{a=1}^6 F_a}= \frac{1}{3}, \, \qquad\qquad \qquad  a=1,\ldots ,6\eeq
Notice that the $SL(4,\mathbb{Z})$ basis  for the Reeb vector in the toric data method is in
general not connected to the analogous basis used in the Molien formula approach.

\section{Appendix: Three dimensional crystals and ``dual ABJM theory''}
\label{app2}

For brane tilings, two of the three complex dimensions of the Calabi-Yau moduli space are
related to the two non-trivial cycles of the two-torus (and there is also a ``radial''
direction). It is therefore natural to associate three dimensional crystals to Calabi-Yau
fourfolds. A simple three dimensional generalization of the hexagonal tiling gives the brane
crystal for $\IC^4$ \cite{Lee:2006hw}.

\begin{figure}[ht]
\begin{center}
  \includegraphics[totalheight=4cm]{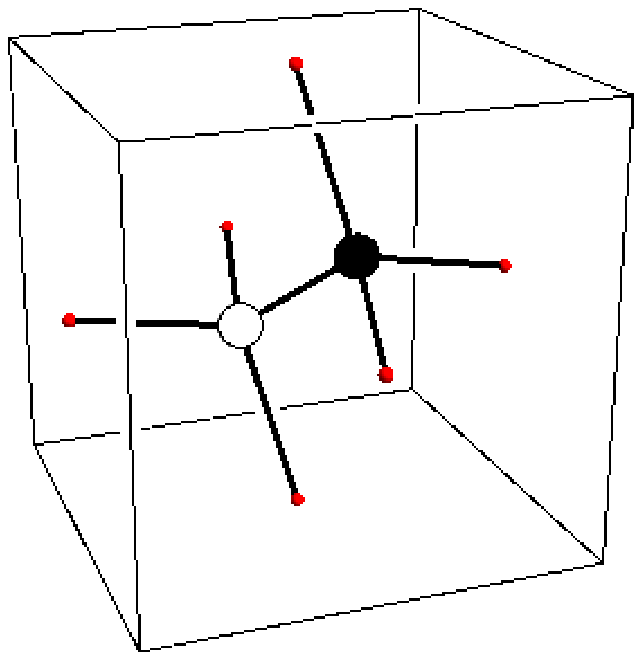}
  \caption{Brane crystal for $\IC^4$. }
  \label{3d_c4}
\end{center}
\end{figure}

Unlike in the case of tilings, in three dimensions it is {\it a priori} unclear how one should
determine the number of gauge groups. The edges correspond to matter multiplets but their
charges cannot be easily read off from the crystal. The order of the fields in the
superpotential is also ambiguous.

\begin{figure}[ht]
\begin{center}
 \hskip -9cm
  \includegraphics[totalheight=7cm]{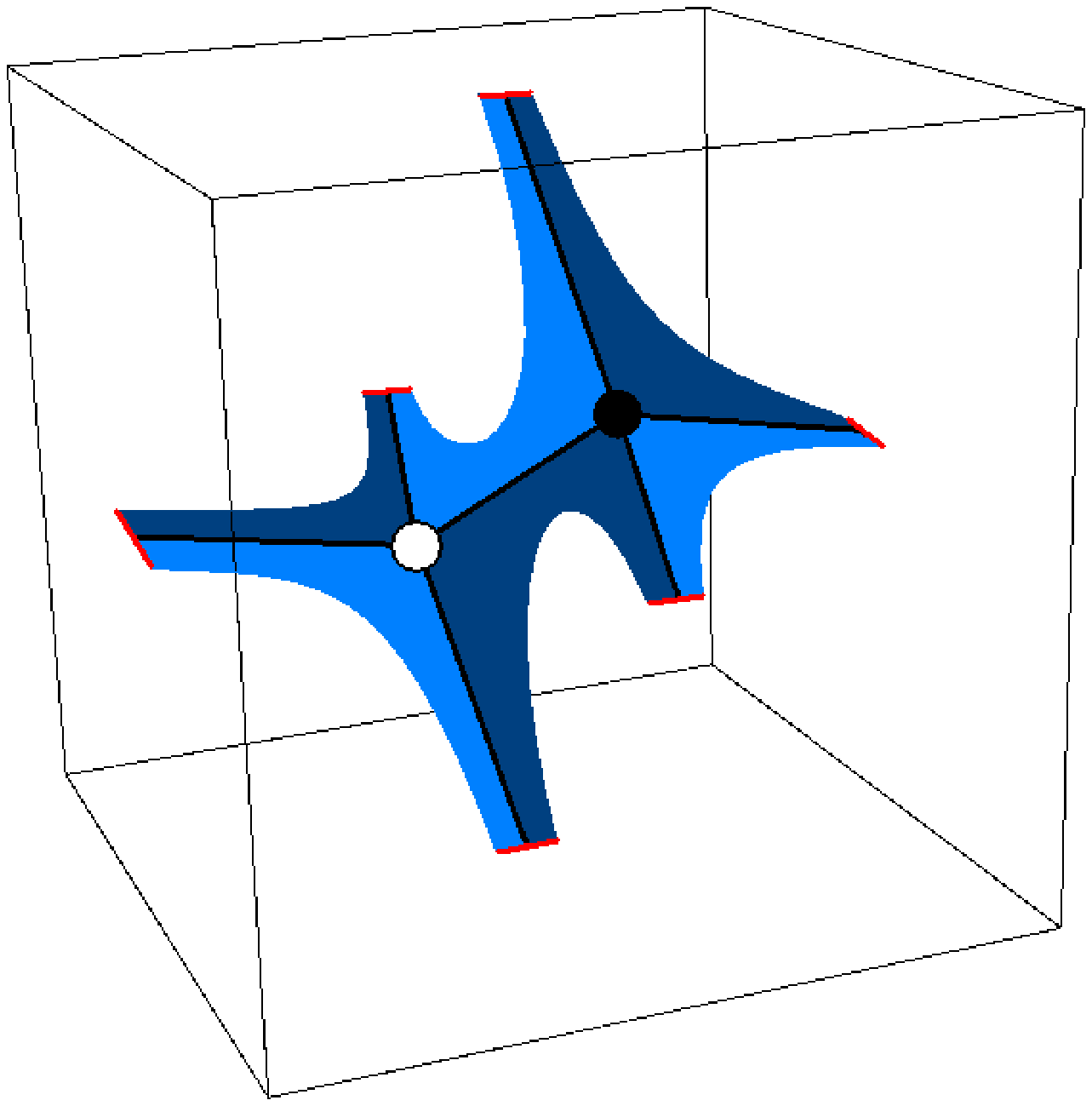}
  \vskip -7cm
  \hskip 9cm
  \includegraphics[totalheight=7cm]{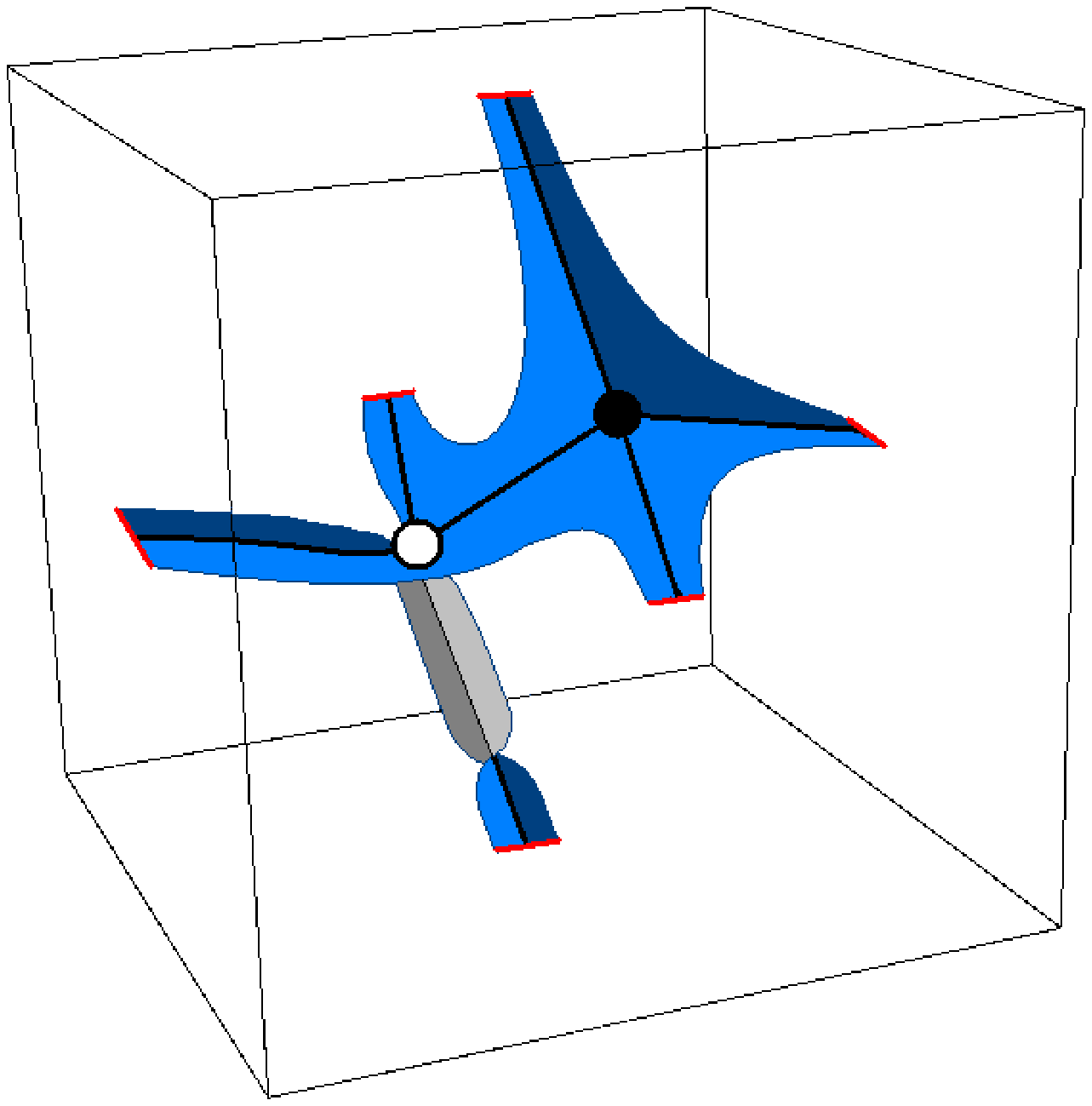}
  \caption{Brane crystal for $\IC^4$ with ribbon graphs: (i) this choice gives the ABJM theory. The red lines indicate where the ribbon touches the
  wall of the fundamental domain. \ (ii) The local plane at the white vertex is now different.
  Gray color indicates the other side of the ribbon. This choice gives the ``dual ABJM theory''. }
  \label{3d_c4_ribbon}
\end{center}
\end{figure}

This latter ambiguity can be fixed by choosing an oriented plane at each vertex. When glued
together, these planes form a ribbon graph which is on top of the brane crystal. The simplest
example is depicted in \fref{3d_c4_ribbon} (i) where the ribbon graph is shown in blue. Once we have
fixed the local plane at the vertices by specifying how the ribbon graph is spanned on the brane
crystal, we can turn to the problem of gauge groups.

A natural choice is to associate the gauge groups to paths on the crystal edges which always
turn left (or always turn right). In two dimensions, this simply gives the faces of the brane
tiling. In three dimensions, however, such paths do not close. They are in fact infinite spirals
as the reader may check in \fref{3d_c4_ribbon}. Each of them defines a ``Burgers vector'': by
moving around a would-be face once, we find ourselves in a different fundamental domain. The
``faces'' spanned by these spirals are infinite half-helicoids. Note that the Burgers vectors
for the faces are parallel (opposite direction).

We see that this definition results in two distinct gauge groups that correspond to the darker
blue and lighter blue areas of the ribbon graph in \fref{3d_c4_ribbon} (i). The matter
multiplets are charged under the gauge groups that correspond to the two sides of the ribbon
near the particular edge. With this definition, we end up with the ABJM theory with four
bifundamental fields between the two gauge groups.

From a certain direction, this three dimensional zinc-blende crystal looks like a square lattice
\fref{coni1}. This is the origin of the similarity of ABJM to the conifold
theory.

So far we have been discussing a particular choice of ribbon graph on the crystal. Another
choice is shown in \fref{3d_c4_ribbon} (ii).  This choice again gives two gauge groups. There
are two bifundamental fields between them and one of the groups has two adjoint fields as well.
The interactions are specified by the superpotential
\be
  W =  [\phi_1 , \phi_2] A B
\ee

\begin{figure}[ht]
\begin{center}
  \includegraphics[totalheight=3.0cm]{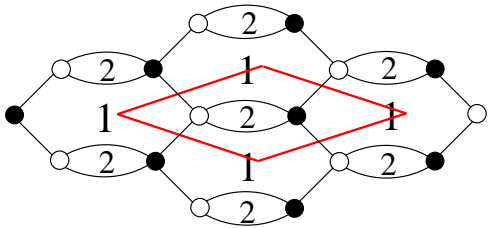}
  \caption{Brane tiling for the ``dual ABJM theory''. }
  \label{dual_abjm}
\end{center}
\end{figure}

We can call this model the ``dual ABJM theory''. An equivalent two-dimensional tiling is shown
in \fref{dual_abjm}. In 3+1 dimensions, this tiling would give an inconsistent theory: some
fields would have vanishing R-charges. In 2+1 dimensions, however, we cannot exclude such models
from the discussion on similar grounds. Such ``inconsistent'' tilings would then greatly enlarge
the set of models describable by two-dimensional tilings.

\begin{figure}[ht]
\begin{center}
  \hskip -5cm
  \includegraphics[totalheight=1.0cm]{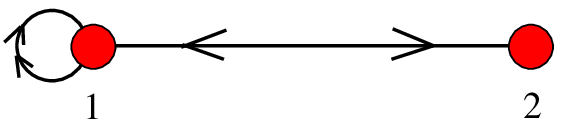}
  \vskip -1.2cm
  \hskip 7cm
  \includegraphics[totalheight=1.5cm]{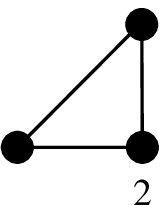}
  \caption{(i) Quiver corresponding to the ``dual ABJM theory''. \ (ii) Toric diagram: one of the external nodes is a double point.}
  \label{dual_toric}
\end{center}
\end{figure}

The Kasteleyn matrix for the ``dual ABJM theory'' is
\be
  K = 1+x+y+z 
\ee

The moduli space at level $k$ is therefore $\IC^2 / \IZ_k \times \IC^2$. At level one, this is
simply $\IC^4$. Thus, we expect the theory to be dual to ABJM in the sense of Seiberg-duality.
This needs further investigation.



In general, one can introduce ``Dehn twists'' for the ribbon graph along the crystal edges. The
possibilities are constrained by the fact that the graph must be oriented otherwise the ordering
of the fields in the superpotential is again ambiguous. Note that there is no clear difference
between the ``tiling'' and the ``untwisted tiling'' (a.k.a. shiver) in three dimensions. For
$\IC^4$, using the high symmetry of the crystal, one can fix the local plane at the black vertex
as in \fref{3d_c4_ribbon} without losing generality. There are then six choices for the local
plane at the white vertex (some of them are mirror images to the ABJM and dual ABJM
ribbon-crystals).


\bibliography{m2}
\bibliographystyle{JHEP}

\end{document}